\documentstyle[12pt,epsfig,cite]{article}
\newcommand{\av}[1]{\mbox{$ \langle #1 \rangle $}}
\newcommand{\lsim}{\raisebox{-0.5mm}{$\stackrel{<}{\scriptstyle{\sim}}$}}
\newcommand{\gsim}{\raisebox{-0.5mm}{$\stackrel{>}{\scriptstyle{\sim}}$}}
\newcommand{\gsp}{\mbox{$\gamma^* p$}}
\newcommand{\gsel}{\mbox{$\gamma^* p \rightarrow \rho p$}}
\newcommand{\gspd}{\mbox{$\gamma^* p \rightarrow \rho Y$}}

\newcommand{\rh}{\mbox{$\rho$}}

\newcommand{\ph}{\mbox{$\phi$}}
\newcommand{\om}{\mbox{$\omega$}}
\newcommand{\jpsi}{\mbox{$J/\psi$}}
\newcommand{\Qsq}{\mbox{$Q^2$}}
\newcommand{\qsq}{\mbox{$Q^2$}}
\newcommand{\W}{\mbox{$W$}}

\newcommand{\y}{\mbox{$y$}}

\newcommand{\bsl}{\mbox{$b$}}

\newcommand{\mpipi}{\mbox{$M_{\pi^+\pi^-}$}}
\newcommand{\mkk}{\mbox{$M_{K^+K^-}$}}

\newcommand{\modt}{\mbox{$|t|$}}

\newcommand{\rzzzz}{\mbox{$r_{00}^{04}$}}
\newcommand{\rzero}{\mbox{$r_{00}^{04}$}}

\newcommand{\gev}{\mbox{\rm GeV}}

\newcommand{\Gevc}{\mbox{\rm GeV/c}}
\newcommand{\GeV}{\mbox{\rm GeV}}

\newcommand{\GeVsq}{\mbox{${\rm GeV}^2$}}
\newcommand{\gevsq}{\mbox{${\rm GeV}^2$}}

\newcommand{\gevcsq}{\mbox{${\rm GeV/c^2}$}}
\newcommand{\Gevcc}{\mbox{${\rm GeV/c^2}$}}
\newcommand{\mevcsq}{\mbox{${\rm MeV/c^2}$}}

\newcommand{\GeVsqm}{\mbox{${\rm GeV}^{-2}$}}
\newcommand{\gevsqm}{\mbox{${\rm GeV}^{-2}$}}

\newcommand{\pbinv}{\mbox{${\rm pb^{-1}}$}}
\def\ar#1#2#3   {{\em Ann. Rev. Nucl. Part. Sci.} {\bf#1} (#2) #3}
\def\err#1#2#3  {{\it Erratum} {\bf#1} (#2) #3}
\def\ib#1#2#3   {{\it ibid.} {\bf#1} (#2) #3}
\def\ijmp#1#2#3 {{\em Int. J. Mod. Phys.} {\bf#1} (#2) #3}
\def\jetp#1#2#3 {{\em JETP Lett.} {\bf#1} (#2) #3}
\def\mpl#1#2#3  {{\em Mod. Phys. Lett.} {\bf#1} (#2) #3}
\def\nim#1#2#3  {{\em Nucl. Instr. Meth.} {\bf#1} (#2) #3}
\def\nc#1#2#3   {{\em Nuovo Cim.} {\bf#1} (#2) #3}
\def\np#1#2#3   {{\em Nucl. Phys.} {\bf#1} (#2) #3}
\def\pl#1#2#3   {{\em Phys. Lett.} {\bf#1} (#2) #3}
\def\prep#1#2#3 {{\em Phys. Rep.} {\bf#1} (#2) #3}
\def\prev#1#2#3 {{\em Phys. Rev.} {\bf#1} (#2) #3}
\def\prl#1#2#3  {{\em Phys. Rev. Lett.} {\bf#1} (#2) #3}
\def\ptp#1#2#3  {{\em Prog. Th. Phys.} {\bf#1} (#2) #3}
\def\rmp#1#2#3  {{\em Rev. Mod. Phys.} {\bf#1} (#2) #3}
\def\rpp#1#2#3  {{\em Rep. Prog. Phys.} {\bf#1} (#2) #3}
\def\sjnp#1#2#3 {{\em Sov. J. Nucl. Phys.} {\bf#1} (#2) #3}
\def\spj#1#2#3  {{\em Sov. Phys. JEPT} {\bf#1} (#2) #3}
\def\zp#1#2#3   {{\em Zeit. Phys.} {\bf#1} (#2) #3}
\begin{document}
\bibliographystyle{unsrt}
\begin{titlepage}
\begin{flushleft}
{\tt DESY 97-082    \hfill    ISSN 0418-9833} \\
{\tt May 1997}                  \\
 
\end{flushleft}
\vspace*{4.cm}
\begin{center}
\begin{Large}
\boldmath
\bf{Proton Dissociative $\rho$ and Elastic $\phi$
\linebreak Electroproduction at HERA \\}
\unboldmath 

\vspace*{2.cm}
H1 Collaboration \\
\end{Large}

\vspace*{1cm}

\end{center}

\vspace*{1cm}

%\newpage

\vspace*{2.cm}
\begin{abstract}

\noindent
The electroproduction of $\rho $ mesons with proton diffractive 
dissociation for \linebreak
$\Qsq > 7\ \GeVsq$ and the elastic electroproduction of
\ph\ mesons for $\qsq \ > 6\ \gevsq$
are studied in $e^+ p$ collisions at 
HERA with the H1 detector, for an integrated luminosity of 2.8 \pbinv .
The dependence of the cross sections on $P_t^2$ and
$Q^2$ is measured, and the vector meson polarisation obtained.
The cross section ratio between proton dissociative and elastic 
production of \rh\ mesons is measured and
discussed in the framework
of the factorisation hypothesis of diffractive vertices.
The ratio of the elastic cross section for \ph\ and \rh\ meson 
production is investigated as a function \linebreak of \qsq.
% Results are compared with previous H1 measurements 
% of elastic $\rho$ meson electroproduction.

\end{abstract}
\end{titlepage} 

%=======================================================================
 
\vfill
\clearpage
%\begin{center}
\begin{sloppypar}
\noindent
%   H1AUTS  Author list by names, no. of authors  379
%           status: 26/03/97   09.19.14
 C.~Adloff$^{35}$,                %WUPP-ST                  Adloff              
 S.~Aid$^{13}$,                   %HAM2-LEFT    8/96        Aid                 
 M.~Anderson$^{23}$,              %MANC-ST  10/95           Anderson            
 V.~Andreev$^{26}$,               %LPI -PD                  Andreev             
 B.~Andrieu$^{29}$,               %ECPL-PD                  Andrieu             
 V.~Arkadov$^{36}$,               %ZEUT-ST    10/96         Arkadov             
 C.~Arndt$^{11}$,                 %DESY-ST   1/96           Arndt               
 I.~Ayyaz$^{30}$,                 %PARI-ST       5/96       Ayyaz               
 A.~Babaev$^{25}$,                %ITEP-PD                  Babaev              
 J.~B\"ahr$^{36}$
% ...
%ZUER-ST                  ZurNedden           
 \\
%\end{center}
\bigskip 
 
\noindent
{\footnotesize{%     H1 Institutes as appearing on publications
% ...
$ ^{11}$ DESY, Hamburg, Germany$^a$ \\
% ...
$ ^{13}$ II. Institut f\"ur Experimentalphysik, Universit\"at Hamburg,
                                     Hamburg, Germany$^ a$  \\
% ...
$ ^{23}$ Physics Department, University of Manchester,
                                          Manchester, UK$^ b$\\
% ...
$ ^{25}$ Institute for Theoretical and Experimental Physics,
                                                 Moscow, Russia \\
 $ ^{26}$ Lebedev Physical Institute, Moscow, Russia$^ f$ \\
% ...
$ ^{29}$ LPNHE, Ecole Polytechnique, IN2P3-CNRS,
                             Palaiseau, France \\
 $ ^{30}$ LPNHE, Universit\'{e}s Paris VI and VII, IN2P3-CNRS,
                              Paris, France \\
% ...
$ ^{35}$ Fachbereich Physik, Bergische Universit\"at Gesamthochschule
               Wuppertal, Wuppertal, Germany$^ a$ \\
 $ ^{36}$ DESY, Institut f\"ur Hochenergiephysik,
                              Zeuthen, Germany$^ a$\\
% ...
$ ^a$ Supported by the Bundesministerium f\"ur Bildung, Wissenschaft,
        Forschung und Technologie, FRG,
        under contract numbers 6AC17P, 6AC47P, 6DO57I, 6HH17P, 6HH27I,
        6HD17I, 6HD27I, 6KI17P, 6MP17I, and 6WT87P \\
 $ ^b$ Supported by the UK Particle Physics and Astronomy Research
       Council, and formerly by the UK Science and Engineering Research
       Council \\
% ...
$ ^f$ Supported by the Deutsche Forschungsgemeinschaft \\
% ...
}}
 
\end{sloppypar}
 
%=======================================================================

\newpage

%=============================================================

%=======================================================================
\section{Introduction} \label{sect:intro}

\noindent
Vector meson production in lepton-proton collisions
is a powerful probe for investigating the nature of diffraction.
At HERA, because of the wide kinematic range accessible in $W$, the 
photon-proton centre of mass energy, and in $Q^2$, the photon virtuality
\footnote {$Q^2$ is the negative square of the four-momentum transfer from the
initial to the final state lepton.},
detailed information on the mechanism of 
the diffractive process can be accumulated. The opportunity to study the
production of vector mesons with different quark contents ($\rho$,$\phi$)
in the elastic and proton dissociation channels adds further to
the information.

Many experimental results on elastic  $\rho$, $\om$, $\ph$ and $\jpsi$ 
meson production by quasi-real photons 
($\Qsq \approx 0$) \cite{help,hjpp,zrhop,zomep,zphip,zjpsip} and 
virtual photons ($\Qsq\ \gsim\ 7\ \GeVsq$) \cite{hele,zrhoe,zphie}
have been obtained by the H1 and ZEUS
experiments. 
Also numerous elastic vector meson production data have been  
reported by fixed target experiments \cite{gp_rho,gp_jpsi,nmcrhophie,emcphie}
at lower $W$, providing information about the 
energy behaviour of vector meson production.
However, little is known about the vector meson proton dissociative process.
In the H1 experiment, the use of the forward detectors (see Sect.2) 
makes it possible
to separate efficiently the proton elastic from the proton
dissociative channels and has led to the first results on proton dissociative 
\jpsi\ photoproduction \cite{hjpp}. 
As yet, no data exist in the high \qsq\ region. 
This contrasts with the situation at high energy proton colliders, 
where the proton diffractive dissociation process $pp \rightarrow pY$
has been widely investigated \cite{cdpp,uafpp,uappp,isrpp}.

Different theoretical models have been proposed in order
to describe diffractive vector meson production. 
In the framework of Regge theory \cite{regge}, which
successfully relates many features of hadronic interactions, 
diffractive vector meson production is described using 
the Vector Dominance Model (VDM) \cite{VDM,Bauer}.
Several QCD models describe diffraction as an exchange of a two gluon
system \cite{Low_Nussinov} adopting either a 
non-perturbative \cite{DL,Cudell} or a perturbative approach. 
In the latter case, either different variants 
of a constituent quark model \cite{Ryskin,Zakha,Haak,Ginzburg} or 
a leading order logarithmic approximation have been used \cite{Brodsky,Levin,F_K_S}. 
These models lead to different predictions for the centre of mass energy 
dependence of the  $\gsp$ cross section. In the non-perturbative 
case it is ``soft'', i.e. similar to that measured in elastic
hadron-hadron scattering, while for the perturbative calculations
a rapid increase of the cross section with $W$ is obtained due to the rise 
of the gluon distribution in the proton in the low Bjorken-$x$ region.
All calculations predict a similar $Q^{-6}$ behaviour for the $\gsp$ 
cross section, which in the perturbative approach may be 
modified to account for the evolution of the parton distributions and 
quark Fermi motion \cite{F_K_S,Nemch}. 
At high $\qsq$ the vector mesons are expected to be mostly 
longitudinally polarised \cite{Cudell}.

The first part of the paper presents the first results on 
$\rho$ meson production with proton dissociation for $\qsq > 7\ \gevsq$.
The cross section ratio for proton dissociative 
to elastic $\rho$ production, which is less sensitive to theoretical 
and experimental uncertainties than absolute cross section values, 
is measured in four intervals of $W$ and $\qsq$.
The $\qsq$ dependence and polarisation are determined for the proton 
dissociation process.
The second part of the paper presents data on elastic \ph\ meson production,
with emphasis on the \qsq\ evolution of the cross section ratio of elastic 
\ph\ to elastic \rh\  production.

%=======================================================================
%=======================================================================
\section{H1 Detector and Event Selection} \label{sect:det}

The data correspond to an integrated luminosity of 2.8 \pbinv. They
were collected in 1994,
when the HERA collider was operated with positrons of 27.5 \gev\
interacting with protons of 820 \gev .
The H1 detector is described in detail in \cite{h1det}. 
As the event selection in both analyses presented have many features 
in common, they will be treated together in this section. Analysis 
specific cuts will be addressed in the appropriate sections.

The basic event topology selected for both analyses consists of a
positron recorded in the backward electromagnetic 
calorimeter (BEMC) and two oppositely charged particles, 
originating from a vertex situated in the nominal $e^+p$ interaction region
which are detected in the tracking system.
The detectors placed in the forward region of the H1 detector
\footnote {In the H1 coordinate system, the direction of the positive $z$ axis 
coincides with the direction of the proton beam, 
defining the ``forward'' region.
The polar angle is defined relative to the $z$ axis.}
are used to distinguish between elastic and proton dissociation events.

The positron is identified as an electromagnetic cluster with an energy 
larger than $12\, \gev$, which is reconstructed in the BEMC and to which a 
hit in the backward multiwire proportional chamber (BPC) is associated. 
The polar angle of the scattered positron $\rm{\theta}$ is determined 
from the position of the BPC hit closest to the BEMC cluster and 
the position of the interaction vertex.
The trigger used to collect the present data  required a total 
deposited energy in the BEMC larger than 8 \gev\  outside a square of
$32 \times 32\ {\rm cm^2}$ around the beam pipe.

Two tracks are required to be detected in the central tracking system,
which are assumed to be pions from the $\rho$ meson decay or kaons from
the $\phi$ meson decay. To be accepted, tracks must be reconstructed 
from at least 5 hits in the drift chambers and have a transverse 
momentum larger than $0.1\, \Gevc$. Except for the positron track
and the tracks related to the vector meson decay products, 
no other tracks in the polar angular range 
of $5^{\rm \circ} < \theta < 170^{\rm \circ}$ are allowed.

The central tracking detector is surrounded by 
the liquid argon calorimeter (LAr).
In the \rh\ analysis, in order to suppress the background,
events including clusters with energies in the LAr or BEMC 
exceeding 0.5 $\gev$ and 1.0 $\gev$ respectively, other than 
those associated with the scattered positron or vector 
meson decay products, are rejected. However clusters 
with a pseudorapidity $\eta > 2.5$ 
\footnote {The pseudorapidity, $\eta$ = $-$ ln tan $\theta /2\ $,
is positive in the forward region; $\theta$ is the polar angle.}
are allowed, as they can be due to particles
originating from the decay of the proton dissociative system.
In the analysis of \ph\ meson elastic production, it is required that 
the energy deposition in the forward part of the LAr with $\eta > 2$ 
should be smaller than $1.0\, \gev$.

Proton dissociation events are tagged by the
three following subdetectors:
the forward part ($\eta > 2.5$) of the LAr calorimeter,
the forward muon detector (FMD) and the proton remnant tagger (PRT).
These subdetectors are sensitive to particles either directly emitted
from the dissociative proton system or rescattered in the beam pipe wall or  
material close to the beam pipe and thus allow the detection 
of primary particles with pseudorapidities up to $\sim 7.5$.
In order to tag proton dissociative events it is required that there be
at least one of the following signals: 
a cluster with energy larger than $0.5 \, \GeV$ in the forward part 
of the LAr calorimeter or two pairs of hits in the FMD or one hit in the PRT. 
For elastic events the absence of any signal in all three 
subdetectors is required.

In the \rh\ analysis, an ``anti-$\phi$'' cut is applied 
to suppress the contamination of $\phi$ mesons:
the invariant mass of the two detected particles, assumed to be kaons, 
is required to be larger than 1.05 $\Gevcc$. 
This cut also reduces the background contribution from $\omega$ mesons.

To achieve the best accuracy in the determination of the kinematical variables
the ``double angle'' method \cite{doan} is applied.
With this method the precisely measured polar angles 
of the scattered positron and of the produced vector meson are  
used to compute the other kinematical variables.

%=======================================================================
%=======================================================================
\section{Electroproduction of {\boldmath \rh} Mesons with Proton  
Dissociation} 

%=======================================================================
\subsection{Diffractive Dissociation and Factorisation} \label{sect:factor}

\begin{figure}[htbp]
\vspace{-1.4cm}
\begin{center}
\epsfig{file=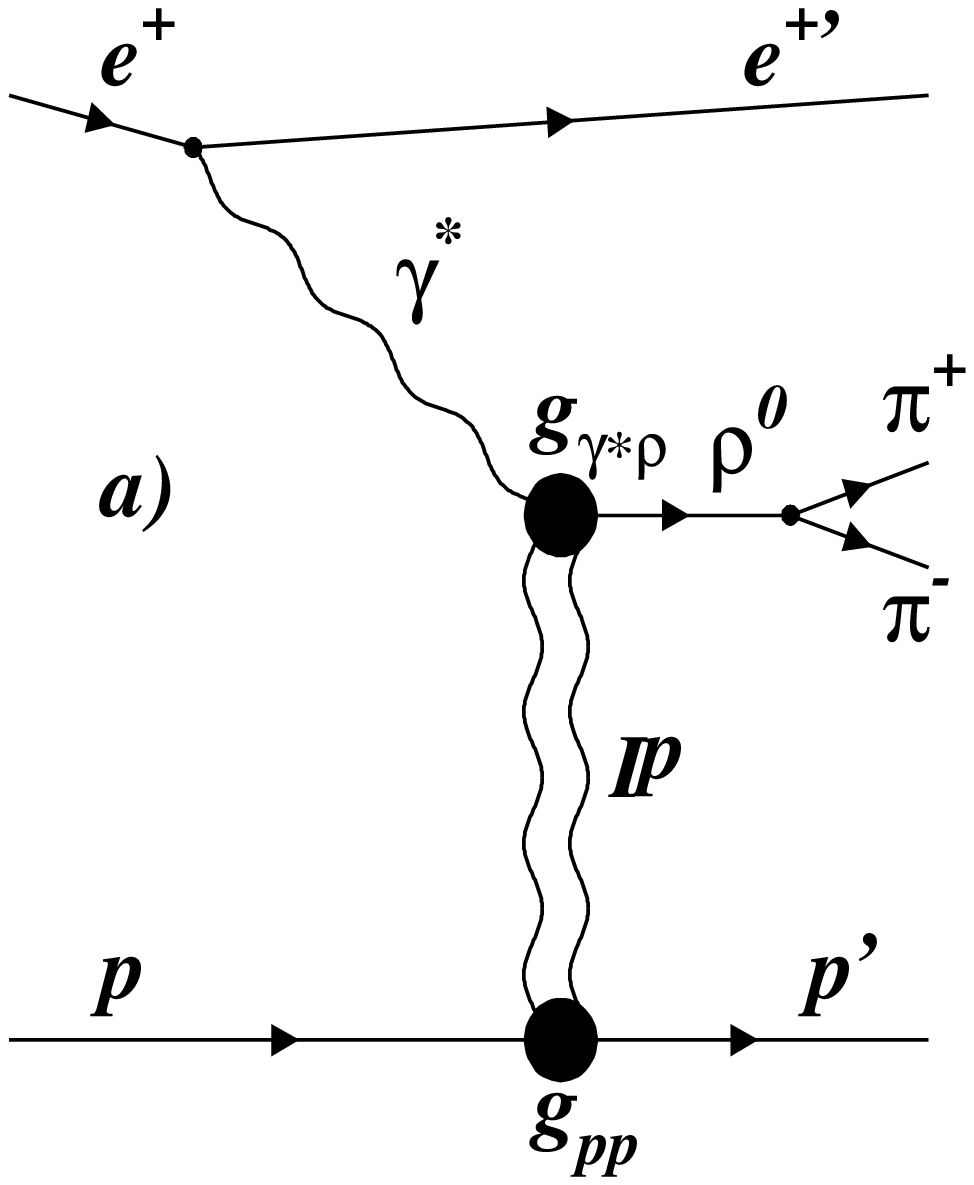,width=8cm,height=8cm}\epsfig{file=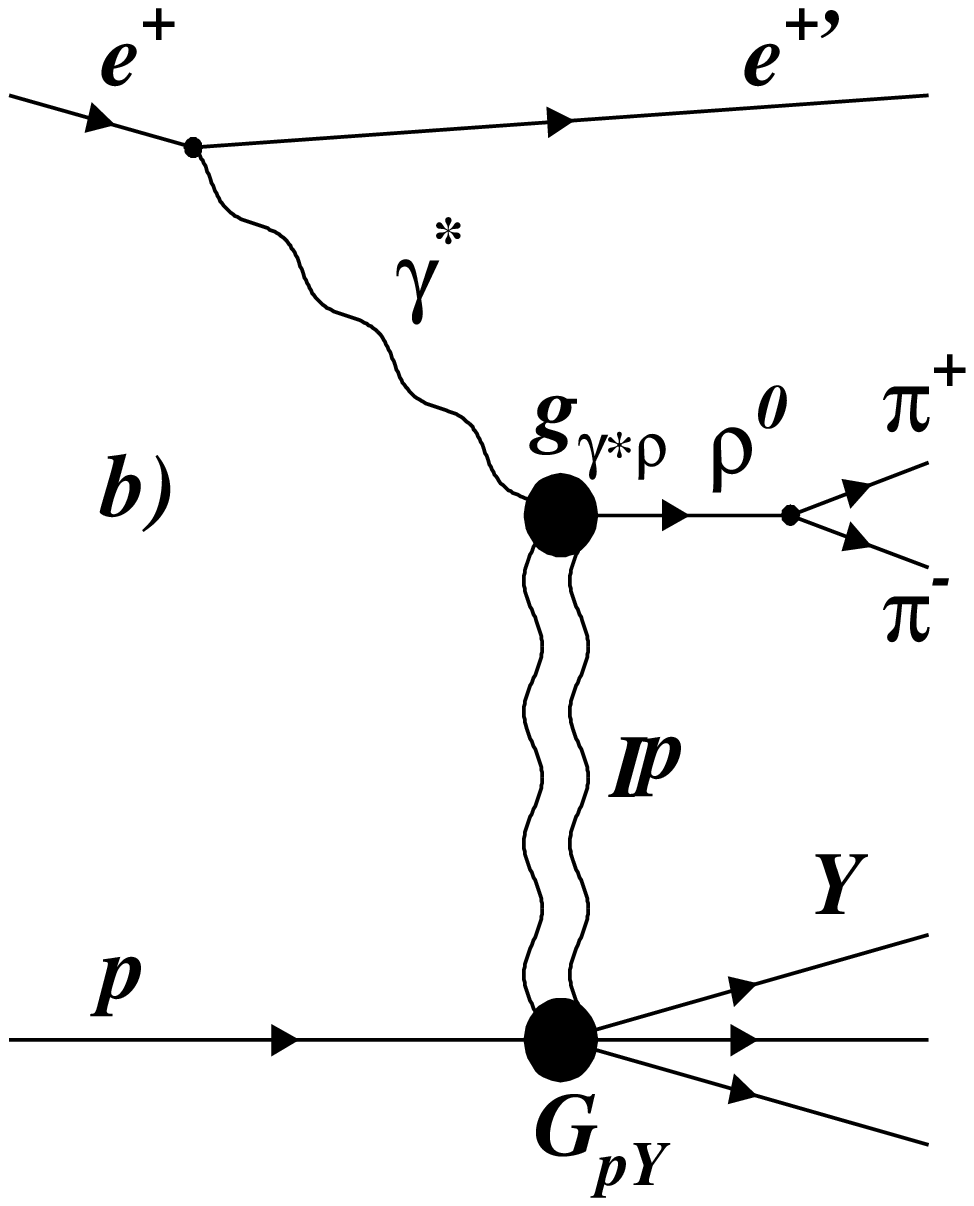,width=8cm,height=8cm}
\end{center}
\vspace{-2.0cm}
\caption{Diffractive $\rho$ meson production: a) elastic scattering;
 b) proton dissociation.}
\label{fig:diag}
\end{figure}

Elastic and proton dissociative \rh\ meson electroproduction 
are illustrated in Fig.\ref{fig:diag}a and \ref{fig:diag}b respectively,
where $Y$ is a low mass system resulting from the proton diffractive
dissociation.

%=======================  \label{fig:diag}  =================================
%=======================  \end{fig:diag}  =================================

The hypothesis of factorisation of the diffractive 
vertex \cite{Kaidalov,Goulianos,factor} naturally appears 
in the framework of Regge theory assuming a single pomeron exchange.
It implies that each amplitude is proportional to the 
product of two vertex functions.
% ; scaling relations 
% between diffractive dissociation and elastic scattering follow.
The differential cross section ratio of proton diffractive 
dissociation to elastic scattering can be expressed 
(see Fig. \ref{fig:diag}) as 
\begin{equation}
\frac{{\rm d}^2 \sigma_{pdis} / {\rm d}t \, {\rm d}M_Y^2 }
{{\rm d} \sigma_{el} / {\rm d}t } \propto
{\left( \frac{\mbox{ g$_{\gamma^* \rho} (t,\Qsq,\lambda)$ G$_{p Y} (t,M_Y)$ }}
{ \mbox{ g$_{\gamma^* \rho} (t,\Qsq,\lambda)$ g$_{p p} (t)$ }} \right) }^2 =
f (t,M_Y) ,
    \label{eq:factor}
\end{equation}
where $M_Y$ is the mass of the proton dissociative system,
$t=(P_{p}-P_{p'(Y)}\,)^2$ the square of the four-momentum transfer 
from the initial to the final state proton (or dissociative system~$Y$), 
$\lambda$ the helicity state of the $\rho$ meson
and g$_{\gamma^* \rho}$, g$_{p p}$ and $G_{p Y}$ 
the vertex functions. The vertex function $G_{p Y}$
can be calculated using a triple-Regge vertex approach \cite{Kaidalov}
for $M_Y$ well in excess of the proton mass.
The vertex function g$_{\gamma^* \rho}$ cancels in eq.~(\ref{eq:factor}) 
and the cross section ratio depends only 
on $t$ and $M_Y$. Hence at fixed $t$, the elastic process and 
the proton dissociative process, independently of $M_Y$, should exhibit 
a similar $\Qsq$ dependence and lead to similar vector meson polarisation.

Other exchange schemes, besides single pomeron exchange,
may lead to a deviation from factorisation.
Deviations due to the contribution of sub-leading reggeon exchanges 
or two-pomeron exchange have been estimated to be of the order 
of $\sim 30 \%$ or less (see \cite{Kaidalov} and references therein).

A basic feature of the diffractive processes is an exponential 
fall ${\rm d} \sigma / {\rm d} t \propto e^{-b |t|} $ of the cross 
section in the low $|t|$ region.
% , where $b$-slope is assumed to reflect 
% the spatial dimensions of the interacting objects.
In the Regge approach the proton dissociative slope of the 
diffractive peak varies as a function of $W$ and $M_Y$ :
\begin{equation}
b(W,M_Y) = b_0 + 2 \alpha ' \ln (W^2/M_Y^2) ,
    \label{eq:slope}
\end{equation}
where the constant $b_0$ can be decomposed as a sum of two
contributions, which define the $t$-dependences of the $\gamma^* \rho$
and triple-pomeron vertices.
The parameter $\alpha '$ is the slope of the pomeron trajectory,
approximately equal to $0.25\, \GeVsqm$ \cite{IDL},
as deduced from hadron-hadron interaction measurements.
The effective $b$-slope of the triple-pomeron vertex
is measured to be approximately $1~\GeVsqm$ \cite{ftest}.
Consequently it is expected that at high $\Qsq$ 
the proton dissociative $b$-slope is considerably smaller then the 
elastic one, the latter reflecting the influence of the size of the proton.
It has to be noted that,
for QCD inspired models, the $b$-slope of the diffractive peak 
at high $\Qsq$ is sometimes assumed to be essentially 
independent of the total energy $W$ \cite{F_K_S},
but this is not a prediction of all models \cite{Zakha}.

In a naive additive quark model the proton dissociative process 
is treated as a quasi-elastic scattering off the constituent quark
in the proton, while the elastic process is treated as 
a coherent scattering off the proton \cite{qmod}.
This approach also predicts a considerable difference between 
the elastic and proton dissociative $b$-slopes.

A detailed experimental study of the proton dissociative 
$pp \rightarrow pY$ process was performed in
several proton collider experiments \cite{cdpp,uafpp,uappp,isrpp}.
The differential cross section ${\rm d} \sigma / {\rm d }M_Y^2$
exhibits an approximate $1/M_Y^2$ dependence. For large
$M_Y$, the slope parameter $b$ is approximately independent of $M_Y$, but
with decreasing $M_Y$ it increases according to eq.~(\ref{eq:slope}) and
even more steeply for very low masses \cite{cdpp}.
The factorisation hypothesis was found to be satisfied to within $\sim 20 \%$
in low energy fixed target experiments \cite{ftest} up to the 
ISR collider energies \cite{isrdd}.
However the proton dissociation cross section rises unexpectedly slowly with 
increasing energy at the SPS and Tevatron colliders, which can be
interpreted as a deviation from factorisation \cite{cdpp,uafpp}.

%=======================================================================
\subsection{Kinematic Selections, Efficiencies and Backgrounds} \label{sect:rhoeff}

In addition to the selection criteria discussed in section~\ref{sect:det},
the kinematic region in this analysis is restricted to:
\begin{equation}
    7 \ < Q^2 <  35 \ \GeVsq , \ \ \
    60 \ <  \  W  \ < \ 180 \ \GeV .
    \label{eq:rhokin}
\end{equation}

The measured reaction is $\gamma^* p\rightarrow \rho Y$, 
where $M_Y^2 / W^2 < 0.05$. Although $M_Y^2$ is not measured
explicitly, the $M_Y^2 / W^2$ range of the measurement is
limited by the forward detector selection criteria. 
The system $Y$ carries most of the momentum of the incoming proton 
and is separated from the $\rho$ meson by a large rapidity gap. 

Such events are modelled by the Monte Carlo (MC) generator,
DIFFVM~\cite{Benno}, which, together with a detailed simulation
of the H1 detector, is used to study efficiencies.
Details of the simulation of the 
dissociated proton system are presented in \cite{hjpp}.
The events are generated with an $M_Y^2$ distribution proportional
to $1/M_Y^{2n}$ with $n$=1.1 for excited masses above 4 $\Gevcc$.
Below 4 $\Gevcc$ the mass distribution is 
taken to follow diffractive dissociation data obtained from 
measurements of proton-deuterium interactions \cite{Goulianos}. 
A systematic uncertainty in the detection efficiency
is estimated by varying the parameter $n$ 
in the $1/M_Y^{2n}$ distribution from $0.9$ to $1.3$ and by 
using different models for the fragmentation of the dissociative 
proton system. This systematic uncertainty is found to be $\sim 7 \%$
and conservatively reflects a model dependence of the forward 
detection efficiency determination for 
the kinematic region $M_Y^2 / W^2 < 0.05$.
A possible proton resonance contribution at 
low $M_Y^2$~\cite{Goulianos} and possible sub-leading exchange 
contributions in the larger $M_Y^2$ region \cite{Kaidalov} are 
encompassed in the above variations of the parameter $n$.
In the very low $M_Y$ region the forward detection 
efficiency decreases with decreasing
$M_Y$, falling to below $50\,\%$ for $M_Y < 1.6\, \Gevcc$.

To estimate the accuracy with which the MC can reproduce the forward 
detector efficiencies, the relative tagging probabilities of proton 
dissociation 
events by the different forward subdetectors obtained by MC are 
compared with those from the data. These comparisons show that the MC 
determination of the forward detection efficiencies has 
an accuracy of $\sim 7 \%$.
The uncertainty arising from statistical fluctuations
in the MC efficiency determination is $\sim 2 \%$.

The background results from the elastic $\rho$ meson
production reaction, $\gamma^* p\rightarrow \rho p$, and 
the  $\gamma^* p\rightarrow \rho Y$ reaction where $M_Y^2 / W^2 > 0.05$.
The latter events survive the selection criteria when at least
one particle with $\eta<2.5$ has not been observed in the detector.
This background is further
suppressed by two additional cuts:
\begin{equation}
     P_t^2  <  0.8 \ \GeVsq , \ \ \
    \sum \ ( E \ - P_z ) > 53 \ \GeV, 
\end{equation}
where $P_t^2$ is the square of the vectorial sum
of the momenta of the electron and pions transverse to the beam direction
\footnote{At HERA energies $P_t^2 \simeq |t|$. For diffractive 
vector meson events, $P_t^2$ is equal to the transverse 
momentum of the scattered proton or dissociative proton system.} 
and the sum is taken over the energy and momentum of the 
pions from the $\rho$ meson decay and the
scattered positron, computed with the ``double angle" method.

Elastic $\rho$ events survive the dissociative selection requirements when
either the elastic proton strikes the beam pipe and is 
detected in the PRT, or when there
is noise in at least one of the forward detectors: the
forward region of the LAr, the FMD or the PRT. The probability
of the proton being detected in the PRT is estimated using the DIFFVM MC
generator and found to be $ 1.0 \pm 0.3 \% $.
The fraction of elastic events with noise in the forward detectors which
would lead to them being identified as proton dissociative is
found by studying events from random triggers. For the forward
region of the LAr the fraction is $ 1.0 \pm 0.2 \% $, for the
FMD $ 3.0 \pm 1.5 \% $ and for the PRT $ \sim 0.1 \% $. 

The non-elastic background can be expected from processes of a
non-diffractive deep inelastic scattering (DIS) or double
(proton and photon) dissociative (DD) nature, which also account
for the non-resonant contribution under the $\rho$ signal.
Such events are studied in the high $P_t^2$ region or using
a sidebands method and are 
found to
have a shallow $P_t^2$ distribution, corresponding to a
$b$-slope value of $0.2 \pm 0.1\, \GeVsqm$.
Therefore, events surviving the
selection cuts but with $P_t^2 > 2.0\, \GeVsq$, where 
the signal event contribution is expected to be negligible, 
are used to estimate the background level. Extrapolating this level to the 
$P_t^2 < 0.8\, \GeVsq$ region, using the $b$ value obtained above, yields an
expected non-elastic contribution of $12 \pm 6 \%$
in the mass interval $0.6~<~M_{\pi^+\pi^-}~<~1.0\,~\Gevcc$.
Included in the $12 \pm 6 \%$ there is a $3 \pm 3 \%$
resonant background part~(i.e. part producing a $\rho$ peak), which is 
obtained from a fit to the mass distribution
of the events at $P_t^2 > 2.0\, \GeVsq$. 
The level of the non-resonant contribution found by this method and 
obtained from the fit of the $\pi^+ \pi^-$ mass 
spectrum~(section~\ref{sect:rhomass})
are in good agreement.

Using the DIFFVM MC, the background contribution from proton dissociative 
electroproduction of $\phi$ and $\omega$ mesons 
is estimated to be smaller than 1$\%$ after applying the ``anti-$\phi$" cut.
The effects of QED photon radiation are
simulated by the HERACLES 4.4 generator \cite{Heracles}
%and the corresponding correction factor is estimated to be $ 0.96 \pm 0.03 $.
and a corresponding correction of $4 \pm 3 \%$ is applied.

%=======================================================================
\subsection{Mass and \boldmath{$P_t^2$} Distributions} \label{sect:rhomass}

\begin{figure}[htbp]
\vspace{-1.1cm}
\begin{center}
\epsfig{file=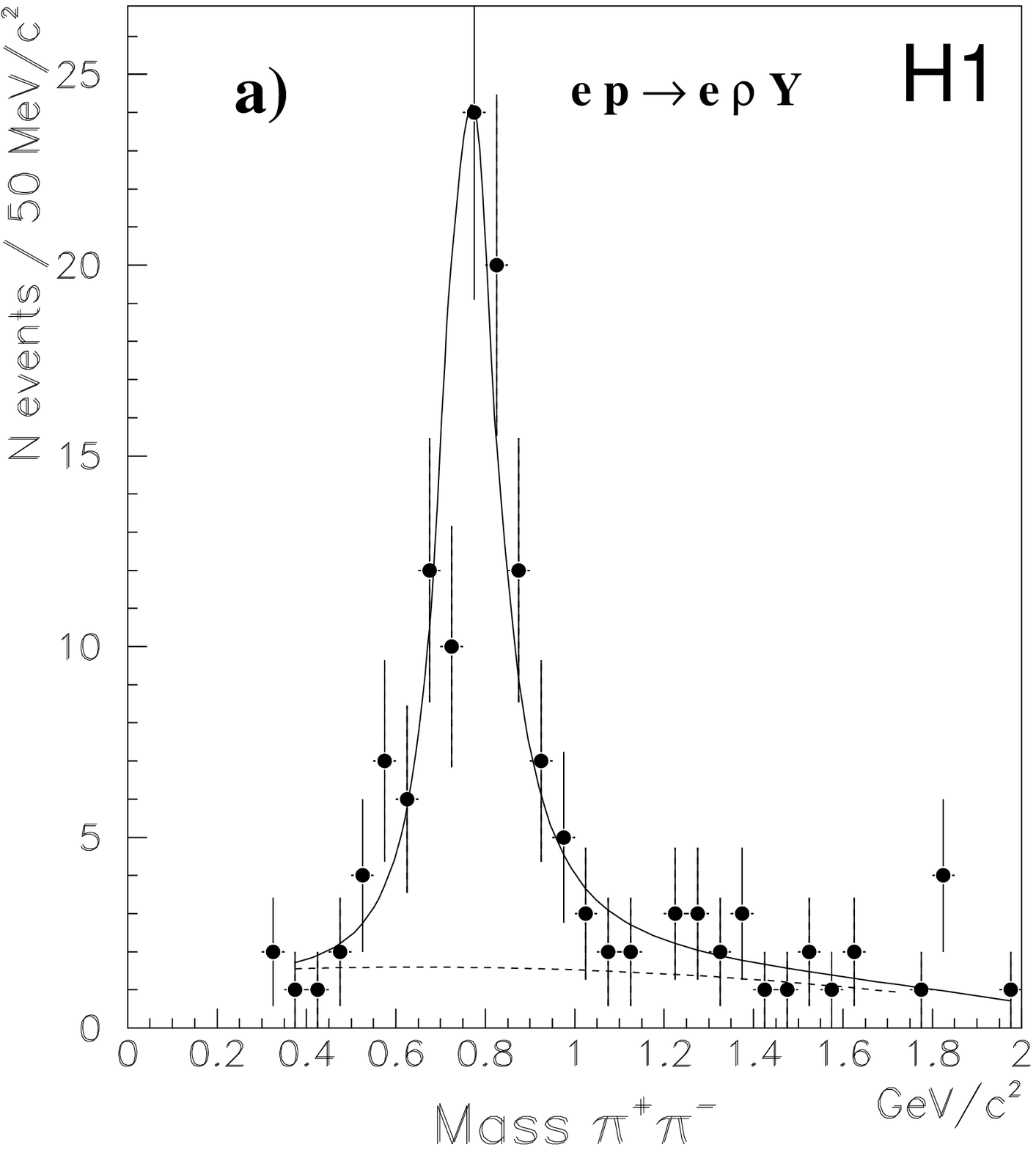,width=8cm,height=8cm}\epsfig{file=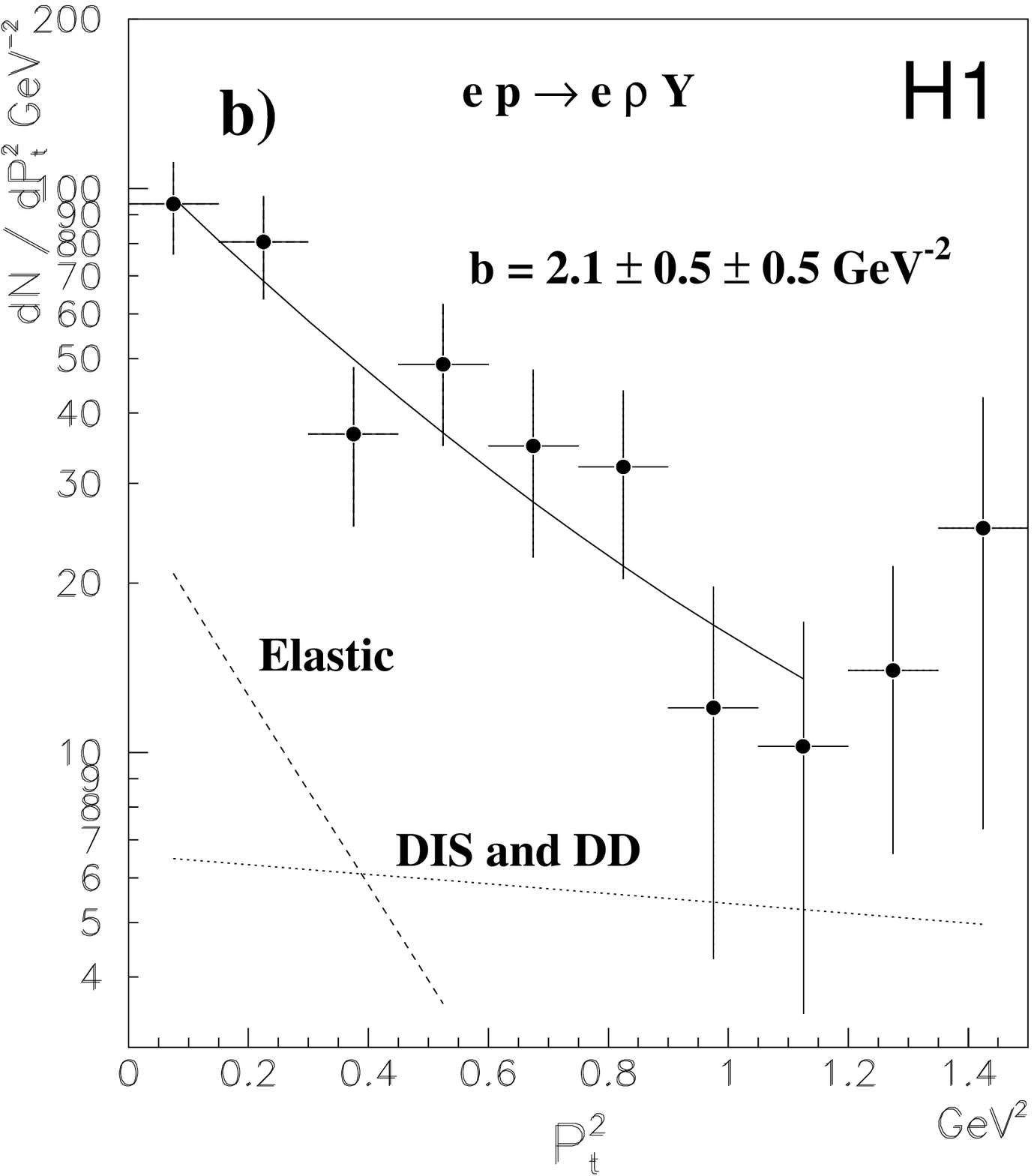,width=8cm,height=8cm}
\end{center}
\vspace{-0.5cm}
\caption{ a) The invariant $\pi^+ \pi^-$ mass spectrum for the
kinematical region $P_t^2 < 0.8\, \GeVsq$. The solid curve 
represents the fit described in the text. The dashed curve shows
the non-resonant background contribution.
b) The efficiency corrected $P_t^2$ distribution for the proton
dissociative data sample. The solid line
represents the fit to the overall distribution, with the elastic (dotted line),
DD and DIS (dashed line) contributions fixed (see text). Only
statistical errors are presented.}
\label{fig:mass}
\end{figure}

The invariant $\pi^+ \pi^-$ mass spectrum of
the proton dissociative data sample is shown in Fig.\,\ref{fig:mass}a for the
kinematical region $P_t^2 < 0.8\, \GeVsq$.
A prominent peak is observed at
the nominal $\rho$ meson mass position. The $\pi^+ \pi^-$ mass
spectrum is fitted with a relativistic Breit-Wigner function 
\cite{Jackson} with a P-wave energy dependent width to describe
the $\rho$ meson signal and a second order polynomial to 
describe the non-resonant background. 
With the $\rho$ mass and width fixed to the nominal 
values \cite{PDG}, the fit gives 101 $\pm$ 13 events corresponding to the 
$\rho$ meson signal.
The systematic error, evaluated by varying the parameterisations of the
signal and background shapes (see \cite{hele}), is found to be $\sim 8 \%$. 
Smearing effects due to detector resolution are estimated using a Monte Carlo 
simulation and the distortion of the $\rho$ meson
signal shape is found to be negligibly small.

%=======================  \label{fig:mass}  =================================
%=======================  \end{fig:mass}  =================================

The acceptance corrected $P_t^2$ distribution is shown in Fig.\,\ref{fig:mass}b
for the selected events for which the $\pi^+ \pi^-$ invariant mass is in the 
interval $0.6 < M_{\pi^+\pi^-} < 1.0\, \Gevcc$.
To extract the exponential slope parameter, the backgrounds are estimated and
subtracted in the fit.
The background from elastically produced \rh\ mesons is evaluated
using the above given MC estimates and taking into account 
the ratio of proton dissociative to elastic events.
An exponential behaviour $e^{-b_{el} P^2_t}$
with $b_{el} = 7.0 \pm 0.8 \pm 0.6\, \GeVsqm$ \cite{hele} has been assumed.
The shape of the $P_t^2$ distribution for the elastic events, in which 
the proton is detected in the PRT detector, is obtained from MC.
The non-resonant $\pi^+ \pi^-$ background is parameterised
with $b_{bg} = 0.2 \pm 0.1\, \GeVsqm$, as outlined above.

With the background contributions fixed, the exponential slope 
parameter $b_{pdis}$
is extracted from a fit to the overall $P_t^2$ spectrum
assuming an exponential $e^{-b_{pdis} P^2_t}$ dependence of the proton 
dissociation cross section. A fit in the region $P_t^2 < 1.2\, \GeVsq$ 
yields a value of $b_{pdis} = 2.1 \pm 0.5\ (stat.) \pm 0.5\ (syst.)\, \GeVsqm$,
where $b_{pdis}$ is the average over all accessible $M_Y$ values.
The systematic uncertainty is due to  the background subtraction
and fit procedure. The uncertainty in the background subtraction is estimated
by varying the background contributions and slopes within errors. 
The uncertainty in the fit procedure is estimated
by varying the bin sizes, the bin positions and $P_t^2$ interval for the fit.

The relatively small $b$-slope value 
for the proton dissociative channel indicates that both 
the $\gamma^* \rho$ and $pY$ vertex functions (see Fig.1b) 
are characterised by small spatial dimensions, in contrast to
the elastic scattering, where the 
$pp$ vertex function corresponds to a value of $\sim 4-5\, \GeVsqm$.
The measured $b$-slope value is close to the value 
obtained in proton dissociative $\jpsi$ photoproduction, where 
$b = 1.6 \pm 0.3 \pm 0.1\, \GeVsqm$ \cite{hjpp}, and to that 
obtained in double dissociative proton-proton scattering,
where the $b$-slope tends to a value of $b_{DD} \sim 1.9\, \GeVsqm$ 
with increasing diffractive masses \cite{isrdd}.
It implies that the contributions of the 
$\gamma^* \rho$, $\gamma \jpsi$ and $pY$ vertex functions to 
the $b$-slopes are of the order of $1\, \GeVsqm$ or less.

The proton dissociative $b$-slope is expected to decrease 
with increasing $M_Y$ according to eq.~(\ref{eq:slope}) and
even faster in the low $M_Y$ region.
Such a behaviour was observed in proton-antiproton collider experiments 
\cite{cdpp}, but has to be tested experimentally for the $\gsp$ process
in the high $Q^2$ region.
To investigate a possible $M_Y$ mass dependence, the event sample is
divided into two subsamples: a high $M_Y$ subsample 
($\av{M_Y} \approx 6.8\, \Gevcc$), in which a cluster is detected in the
forward part of the LAr calorimeter, and a low $M_Y$ subsample 
($\av{M_Y} \approx 2.9\, \Gevcc$), characterised by the absence
of clusters in the forward part of the LAr calorimeter.
The values obtained for the $b$-slopes of the two event samples
are $b_{high} = 2.7 \pm 1.3 \pm 0.7\, \GeVsqm$
and $b_{low} = 1.8 \pm 0.6 \pm 0.6\, \GeVsqm$, respectively.
No evidence for any $M_Y$ dependence of the $b$-slope is thus 
observed beyond the uncertainties of the measurement.
It should be noted that low mass $M_Y\, < \, 1.6\, \Gevcc$ proton 
dissociative events cannot be efficiently tagged by the forward detectors.

%=======================================================================
\subsection{Proton Dissociative to Elastic Cross Section Ratio} \label{sect:rhoratio}

The ratio of the proton dissociative to the elastic $\rho$ production 
cross section is calculated
using the fit results obtained as described above and restricted 
to the kinematic region $7 < Q^2 <  35$ \GeVsq\ and 
$60 < W < 180$ \GeV. The proton dissociative cross section corresponds 
to the region $M_Y^2 / W^2 < 0.05$. Efficiency and 
acceptance corrections for the two event samples, the proton dissociative 
and elastic $\rho$ events, are given in Table \ref{tab:backgr}.

% Table 1
\begin{table}[htbp]
\begin{center}
\begin{tabular}{|l|c|c|c|}
% ----------------------------
\hline \hline
 & Proton & Elastic & Ratio, \\ 
 & dissociation & scattering & corrections \\  \hline \hline
Number of events & 101 $\pm$ 13 & 291 $\pm$ 23 & 0.35 $\pm$ 0.05 \\ \hline \hline
Trigger efficiencies & 0.99 $\pm$ 0.01 & 0.99 $\pm$ 0.01 & 1.00  \\ \hline
Selection efficiencies & 0.39 $\pm$ 0.04 & 0.57 $\pm$ 0.02 & 1.46 $\pm$ 0.15 \\
\hline \hline
$P_t^2$ acceptance correction         & 1.23 $\pm$ 0.15 &
1.00 $\pm$ 0.01                    & 1.23 $\pm$ 0.15 \\ \hline 
Forward detectors off              & 1.10 $\pm$ 0.01 & 0.97 $\pm$ 0.01 & 1.13 $\pm$ 0.02 \\ \hline
Fit procedure                      & 1.00 $\pm$ 0.08 & 1.00 $\pm$ 0.08 & 1.00 $\pm$ 0.05 \\ \hline
Radiative corrections              & 0.96 $\pm $ 0.03 & 0.96 $\pm$ 0.03 & 1.00  \\ \hline
Total background correction (a+b+c)   & 0.82 $\pm$ 0.07 & 0.89 $\pm$ 0.08 & 0.92 $\pm$ 0.11 \\ \hline \hline
a) Elastic background             & 0.15 $\pm$ 0.06 & &   \\ \hline
b) Proton dissociation background & & 0.09 $\pm$ 0.08 & \\ \hline
% DD and DIS bg. corrections       & 0.88 $\pm$ 0.06
% & 0.89 $\pm$ 0.06 &  \\ \hline
c) DD and DIS resonant background & 0.03 $\pm$ 0.03 & 0.02 $\pm$ 0.01 & \\ \hline \hline
% ----------------------------
\end{tabular}
\end{center}
\caption{Numbers of events, efficiencies and correction factors for the 
proton diffractive dissociation and the elastic scattering data samples.}
\label{tab:backgr}
\end{table}

The background contributions were discussed in section 
\ref{sect:rhoeff} and in \cite{hele}.
The entry in Table \ref{tab:backgr}  ``Forward detectors off'' corrects for 
some data taking periods, during which the FMD or PRT subdetectors 
were not operational.
% No uncertainty in the luminosity measurement is included 
% as it cancels in the cross section ratio.
A systematic uncertainty in the luminosity measurement of $\sim 2\%$
cancels in the cross section ratio.

The cross section ratio is corrected for the $P_t^2$ acceptances using 
the exponential $P_t^2$ dependences and taking into account the uncertainty
of the slope measurement. 
Some of the systematic uncertainties in the two event samples are not 
independent 
and therefore totally or partially cancel in the cross section ratio, 
as can be seen from Table \ref{tab:backgr}.

The ratio of the proton dissociative to the elastic $\rho$ meson 
production cross section is measured to be:
\begin{equation}
\frac{\sigma ( e p \rightarrow e \rho Y )}
{\sigma ( e p \rightarrow e \rho p )}=0.65\pm 0.11(stat.)\pm 0.13(syst.).
                                            \label{eq:ratio}
\end{equation}
The systematic error is dominated by the acceptance determinations,
the background estimates and the fit procedure.

The $\rho$ meson electroproduction cross section
is converted into a $\gsp\ $ cross section using the relation
\begin{equation}
\sigma (\gamma^*p \rightarrow \rho Y) = \
\frac{1}{\Gamma_{T}} \frac{{\rm d}^2 \sigma 
(ep \rightarrow e \rho Y)}{{\rm d}W \ {\rm d}Q^2} \ ,
\ \ \Gamma_{T} = \frac {\alpha_{em} \ (1 + (1-\y)^2)} {\pi\  \W \  \Qsq} \ ,
                                            \label{eq:sigma}
\end{equation}
where $y$ is the Bjorken inelasticity variable, 
$\sigma (\gamma^*p \rightarrow \rho Y)$ the virtual photon-proton
cross section and $\Gamma_{T}$ the transverse virtual photon flux factor.
The error on the $\Gamma_{T}$ factor is estimated by varying the $W$ or $\Qsq$
distributions within errors, leading to an 
uncertainty in $\Gamma_{T}$ of $ \sim 7\% $.
To study any dependence of the ratio on $\qsq$ and $W$, the 
data are divided in four ($\qsq$, $W$) bins.
The proton dissociative $\sigma (\gspd)$ cross sections 
and ratios of the proton dissociative to the elastic cross
sections for four kinematic intervals are presented in Table \ref{tab:ratios}.
Within uncertainties, no dependence of the cross section 
ratio on $W$ or $\Qsq$ is observed.

% Table 2
\begin{table}[htbp]
\begin{center}
\begin{tabular}{|c|c|c|}
% ----------------------------
\hline \hline
  & \multicolumn{2}{|c|}{$7 < Q^2 <  35\, \GeVsq$} \\ \hline
  & \multicolumn{2}{|c|}{$60 < W < 180$ GeV} \\ \hline
$\frac{\sigma (\gspd)}{\sigma (\gsel)}$
& \multicolumn{2}{|c|} { 0.65 $\pm$ 0.11 $\pm$ 0.13 } 
\\ \hline \hline
  & \multicolumn{2}{|c|}{$7 < Q^2 <  15\, \GeVsq$} \\ \hline
  & $60 < W < 120$ GeV & $120 < W < 180$ GeV \\ \hline
$\sigma (\gspd)$ (nb) & 24 $\pm$ 5 $\pm$ 6 
& 19 $\pm$ 6 $\pm$ 5 \\ \hline
% Cross section of $\gsel$ (nb) & 30.1 $\pm$ 3.0 & 28.6 $\pm$ 4.0 \\ \hline
$\frac{\sigma (\gspd)}{\sigma (\gsel)}$ &  0.74 $\pm$ 0.17 $\pm$ 0.16 
& 0.63 $\pm$ 0.22 $\pm$ 0.14
\\ \hline \hline
  & \multicolumn{2}{|c|}{$15 < Q^2 <  35\, \GeVsq$} \\ \hline
  & $60 < W < 120$ GeV & $120 < W < 180$ GeV \\ \hline
$\sigma (\gspd)$ (nb) & 3.0 $\pm$ 1.1 $\pm$ 0.8 
& 3.0 $\pm$ 1.6 $\pm$ 0.8 \\ \hline
% Cross section of $\gsel$ (nb) & 4.9 $\pm$ 0.9 & 5.6 $\pm$ 1.3 \\ \hline
$\frac{\sigma (\gspd)}{\sigma (\gsel)}$ &  0.60 $\pm$ 0.26 $\pm$ 0.14 
& 0.51 $\pm$ 0.31 $\pm$ 0.13
\\ \hline \hline
% ----------------------------
\end{tabular}
\end{center}
\caption{The proton dissociative $\sigma (\gspd)$ cross sections 
and ratios of the proton dissociative to the elastic cross
sections for four kinematic intervals.}
\label{tab:ratios}
\end{table}

%=======================================================================
\subsection{\boldmath{$\Qsq$} Dependence and Polarisation } \label{sect:rhoq2}

\begin{figure}[htbp]
\vspace{-1.1cm}
\begin{center}
\epsfig{file=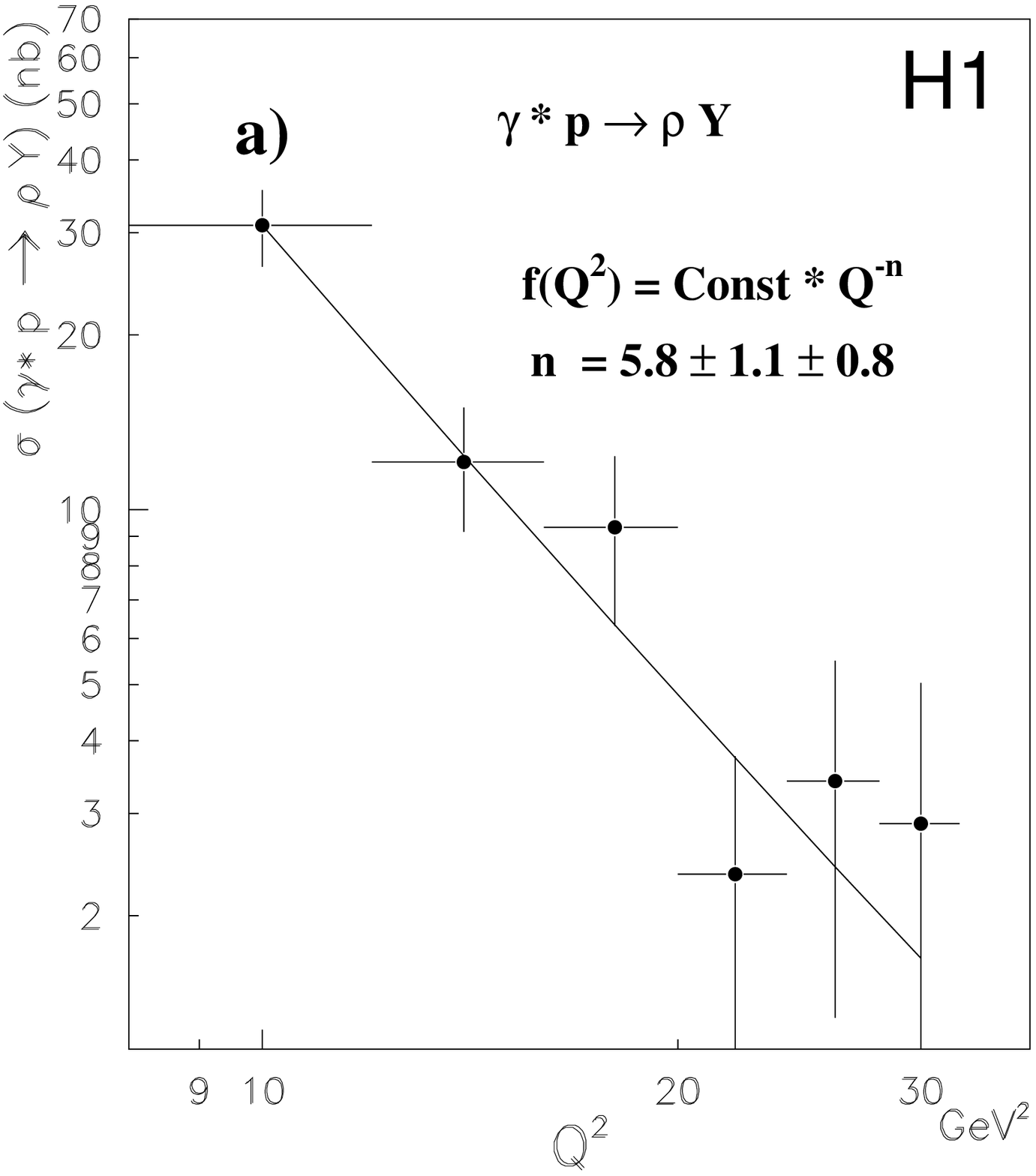,width=8cm,height=8cm}\epsfig{file=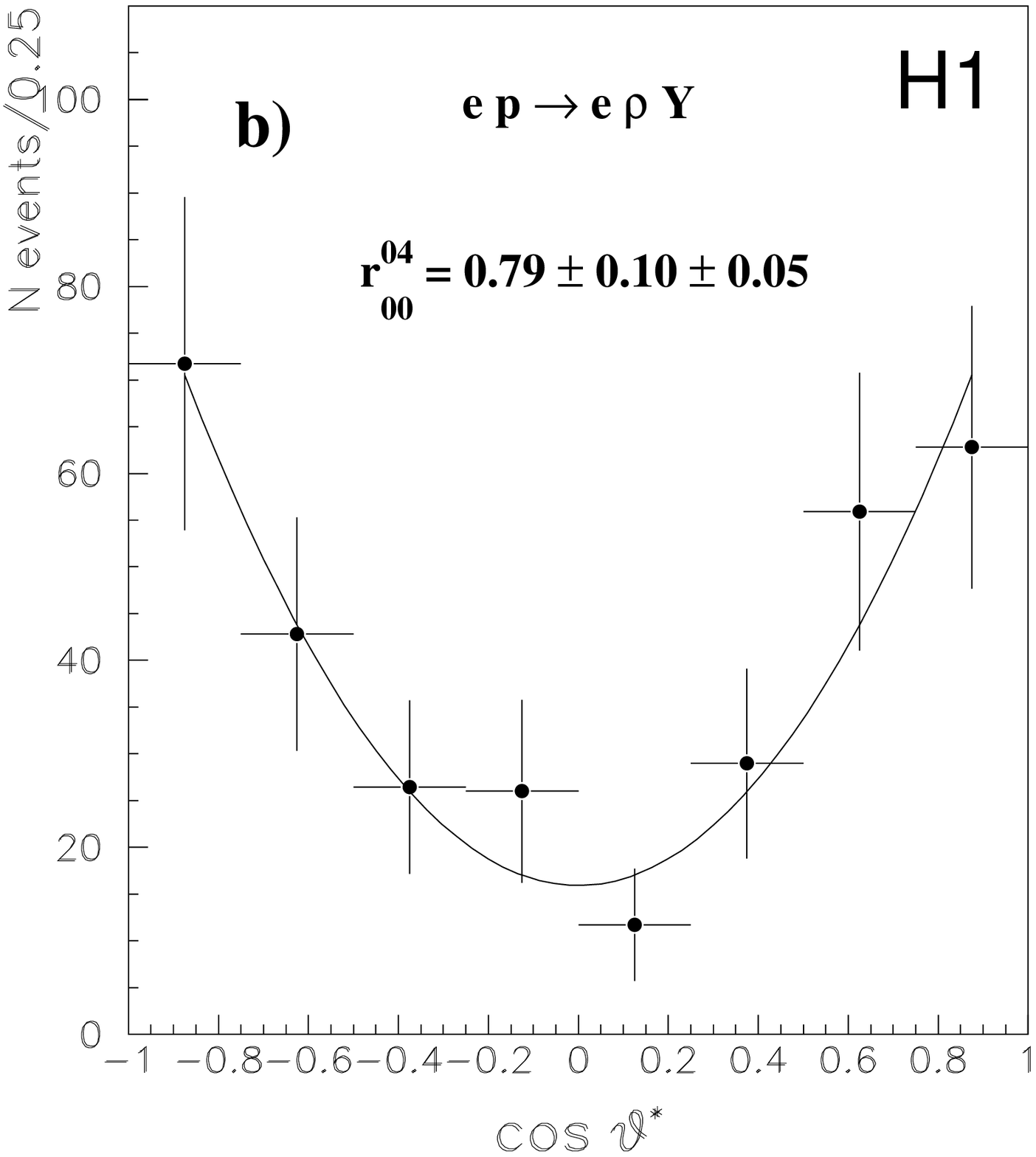,width=8cm,height=8cm}
\end{center}
\vspace{-0.4cm}
\caption{a) $\Qsq$ distribution of the proton 
dissociative $\rho$ meson production $\gsp$ cross section
after efficiency correction. The solid curve
represents a fit with the $Q^{-n}$ function, the backgrounds being taken
into account in the fit procedure.
b) Efficiency corrected $\cos\vartheta^*$ distribution, 
the backgrounds being subtracted in the fit. Only statistical
errors are presented for both distributions.}
\label{fig:qqpol}
\end{figure}

The proton dissociative data in the region 
\mbox{$0.6 < M_{\pi^+\pi^-} < 1.0\, \Gevcc$} are used
for the measurements of the $\gsp$ cross section $\Qsq$ dependence
and $\rho$ polarisation.
The background in the $\Qsq$ distributions from elastic and from  
double dissociative and non-diffractive deep inelastic scattering 
are assumed to have a $Q^{-n}$ behaviour with $n$ values respectively 
\mbox{$n = 5.0 \pm 1.0 \pm 0.4$} (see \cite{hele}) and $n = 3.0 \pm 0.5$,
as obtained from a dedicated background study.
The background subtracted $\Qsq$ distribution for the 
reaction $\gsp \rightarrow \rho Y$ (Fig. \ref{fig:qqpol}a) is well 
fitted  by a $Q^{-n}$ dependence with $ n = 5.8 \pm 1.1 \pm 0.8 $. 
The systematic error comes mainly from the uncertainties in the background 
subtraction, the fit procedure and the photon flux estimate.
The $\Qsq$ dependence is similar to that measured in  
elastic electroproduction \cite{hele,zrhoe}.

%=======================  \label{fig:qqpol}  =================================
%=======================  \end{fig:qqpol}  =================================

Information about the $\rho$ meson polarisation can be extracted 
from the $\cos\vartheta^*$ angular distribution, where
$\vartheta^*$ is the angle in the $\rho$ meson rest frame
between the direction of the $\pi^+$ and the direction of the $\rho$
meson in the $\gamma^* p$ center of mass system. This
distribution is expected to be different for the different 
helicity states : $\propto \cos^2 \vartheta^*$ for longitudinally 
polarised $\rho$ mesons and $\propto \sin^2 \vartheta^*$ for
transversely polarised $\rho$ mesons.
% \cite{Schilling_Wolf}.
Explicitly the angular distribution can be expressed in terms
of the appropriate $\rho$ meson spin density matrix element \rzzzz\ as: 
\begin{equation}
\frac {{\rm d}N} {{\rm d}\cos\vartheta^*} 
\propto  1 - \rzzzz + (3 \ \rzzzz -1) \ \cos^2\vartheta^*  ,
                  \label{eq:costhst}
\end{equation}
where $r^{04}_{00}$ is the probability
for the $\rho$ meson to be longitudinally polarised.

The acceptance corrected $\cos \vartheta^*$ angular distribution is 
shown in Fig. \ref{fig:qqpol}b.
The backgrounds are subtracted in the fit procedure.
A flat distribution is assumed for DD and DIS and the value
$r^{04}_{00}= 0.73 \pm 0.05 \pm 0.02$ (see \cite{hele}) 
is used for the elastic scattering background subtraction. 
The fit yields $r^{04}_{00} = 0.79 \pm 0.10 \pm 0.05$, again similar to
the elastic scattering data value \cite{hele} and
indicating that the $\rho$ mesons are mostly longitudinally polarised
also for the proton dissociative process. 
The systematic uncertainties are mostly due to the uncertainties of the 
background subtraction and the fit procedure.

%=======================================================================
\subsection{Test of the Factorisation Hypothesis} \label{sect:rhotest}

The results obtained for $\gsp$ elastic scattering can be
compared with other diffractive processes in the framework 
of the factorisation hypothesis using eq.~(\ref{eq:factor}).
For example, in the case of proton-proton collisions the coupling constant 
g$_{\gamma^* \rho}$ is replaced by g$_{pp}$,
which cancels in the ratio.

Assuming an exponential $\modt$ dependence of the elastic 
and proton dissociative processes \newpage \noindent
and a universal $M_Y$ dependence of proton dissociation
\footnote {Both assumptions are in a good agreement with the 
experimental data for the low $|t|$ region and possible differences 
between the $M_Y$ dependences of the $pp$ and $\gsp$ 
processes due to the variation of the $b$-slopes with $M_Y$ are expected 
to be inside the theoretical and experimental uncertainties.},
eq.~(\ref{eq:factor}) leads after integration
over a fixed $M_Y$ interval to:

\begin{equation}
\frac{{\rm d}\sigma_{pdis}/{\rm d}t(t=0) }{{\rm d}\sigma_{el}/{\rm d}t(t=0) } =
\frac{\sigma_{pdis} b_{pdis} }{\sigma_{el} b_{el} } = f_1 , \ \ \ \ \ \ \ \
b_{el}- b_{pdis} = f_2 ,
                                            \label{eq:test}
\end{equation}
where $\sigma_{el}$ and $\sigma_{pdis}$ are the elastic and
proton dissociative cross sections integrated over $t$ and $M_Y$.

If factorisation holds, $f_1$ and $f_2$  are expected 
to be the same in different diffractive processes. 
Table 3 compares the values for $f_1$ and $f_2$ in $\gamma^{(*)} p$ 
and $pp$ interactions, albeit with somewhat different 
centre of mass energies. Such energy variations are not expected
to affect this comparison.
Note that after efficiency corrections similar
$M_Y^2 / W^2 < 0.05$ regions are used in the measurements.

% Table 3
\begin{table}[htbp]
\begin{center}
\begin{tabular}{|l|c|c|c|}
% ----------------------------
\hline \hline
Experiment & ISR, $pp \to pY$ \cite{isrpp} & H1, $\gsp \to \rho Y$ & 
H1, $\gamma p \to \jpsi Y$ \cite{hjpp} \\ \hline
cms energy, GeV & 53 & 60 - 180 & 30 - 150  \\  \hline \hline
$b_{el}$, GeV$^{-2}$ & 13.1 $\pm$ 0.3 & 7.0 $\pm$ 1.0 & 4.0 $\pm$ 0.3 \\ \hline
$b_{pdis}$, GeV$^{-2}$ & 6.5 $\pm$ 1.0 & 2.1 $\pm$ 0.7 & 
1.6 $\pm$ 0.3 \\ \hline
$\sigma_{pdis} /\sigma_{el}$ & 0.48 $\pm$ 0.03 & 0.66 $\pm$ 0.17 & 
1.0 $\pm$ 0.2 \\ \hline
$f_1$ & 0.24 $\pm$ 0.04 & 0.20 $\pm$ 0.09 & 0.40 $\pm$ 0.11 \\ \hline
$f_2$, GeV$^{-2}$ & 6.6 $\pm$ 1.0 & 4.9 $\pm$ 1.2 &
2.4 $\pm$ 0.4 \\ \hline \hline
% ----------------------------
\end{tabular}
\end{center}
\caption{Comparison of the $f_1$ and $f_2$ values 
for $\gamma^{(*)} p$ and $pp$ collisions.}
\label{tab:factor}
\end{table}

It is observed that the $\rho$ meson data with $Q^2 > 7$ \GeVsq\ 
are consistent within errors with the proton-proton results
from the ISR \cite{isrpp}, whereas the $f_1$ and $f_2$ values obtained 
in $\jpsi$ photoproduction \cite{hjpp} are somewhat different.

%==============================================================================
\section{Elastic Electroproduction of {\boldmath \ph} Mesons}
%
%==============================================================================
\subsection{Energy Dependence of Vector Meson Production}

Contrasting results are obtained for the $W$ dependence of the 
cross sections for elastic production of different vector mesons and at
different photon virtualities.
On the one hand, high energy \rh\ meson production 
by quasi-real photons exhibits 
behaviour typical of hadronic interactions, in particular a slow increase 
of the cross section with energy: 
$\sigma\ (\gamma p \rightarrow \rho p) \propto (W^2)^{2 \epsilon}$, where 
$\epsilon \approx 0.08$ (at $|t| = 0\ \gevsq$)\cite{DoLa}, 
and shrinkage of the diffraction peak, i.e. an increase of the \bsl\ 
slope with increasing energy.
It is therefore attributed essentially to soft pomeron exchange, dominated by 
QCD non-perturbative features.
On the other hand, the cross section for photoproduction of \jpsi\  mesons 
\cite{hjpp, zjpsip} 
increases much faster with energy ($\epsilon \approx 0.25$), as
qualitatively expected in models in which the large mass of the charm quark
provides a hard scale for perturbative QCD 
and the pomeron is interpreted
as a two gluon exchange.
At higher photon virtualities, $\qsq\ \gsim\ 8\ \gevsq$, measurements of
\rh\ production \cite{hele,zrhoe} indicate a steeper rise of the 
cross section with energy, indicative of a transition regime between soft 
and hard processes, whereas the behaviour of the \jpsi\ cross section is 
similar to that of photoproduction \cite{hele}.

It is thus interesting to study the 
electroproduction of \ph\ mesons, which have a mass between those
of the \rh\ and \jpsi\ mesons, and presumably a more compact wave function
than the \rh\ meson. Moreover, \ph\ meson production 
is ``OZI'' exotic \cite{ozi} in both the $s$ and 
$t$ channels and is thus dominated by pomeron exchange.

The photoproduction of \ph\ mesons \cite{zphip,Bauer,fixedtgtphiphot} 
exhibits features typical of soft diffractive interactions, similar to 
the \rh\ case. It is of particular interest to study
the $\ph / \rh$ cross section ratio evolution with \qsq\ and \W. 
This ratio is measured in photoproduction to be well below the 
value $2 / 9$ expected from quark charge counting and SU(3) symmetry. 
Both in the soft pomeron and in the perturbative QCD 
approaches \cite{F_K_S, Nemch} the ratio increases with \qsq, and in the 
latter case exceeds the value $2 / 9$ at high \qsq.

The only existing results on \ph\ meson electroproduction for 
$\qsq\ \gsim\ 6$ \gevsq\ are from the EMC \cite{emcphie} and NMC \cite{nmcrhophie} 
Collaborations at $\av{W} \approx 14$ \gev\ and the ZEUS Collaboration
\cite{zphie} at HERA \footnote{The electroproduction of \ph\ mesons at
$\qsq\ \lsim\ 2.5$ \gevsq\ has also been measured in fixed target experiments
for $W < 4$ \gev\ \cite{cassel} and for $W \approx 12$ \gev\
\cite{shambroom}.}. 
%We report here \cite{Eilat} on the data obtained in 1994 by the H1 Collaboration,
%when HERA operated with 820 GeV protons and 27.5 GeV positrons. 
%They correspond to an integrated luminosity of 2.5 \pbinv.  

%=%==============================================================================
%
\subsection{{\boldmath $ep$ and \gsp} Cross Sections}
 The production of \ph\ mesons is studied in the kinematic region

\begin{equation}
    6 < \qsq < 20\ \gevsq ,  \ \ \
    0.02 < y < 0.2 \ .  
    \label{eq:phikin}
\end{equation}
These cuts are equivalent to $42 < W < 134$ \gev .
In addition to the selection criteria presented in section \ref{sect:det},
the following cuts are applied to select the elastic \ph\ sample:

\begin{equation}
    P_t^2 \ < 0.6\ \gevsq , \ \ \ 
    \sum \ ( E \ - P_z ) > 45 \ \GeV \ ,
    \label{eq:phi_pt}
\end{equation}
where the sum runs over the measured energy and momentum of the
kaons from the \ph\ meson decay and the 
scattered positron. These cuts suppress non-diffractive backgrounds and 
reduce radiative corrections respectively.

\begin{figure}[htbp]
\vspace{-1.1cm}
\begin{center}
\epsfig{file=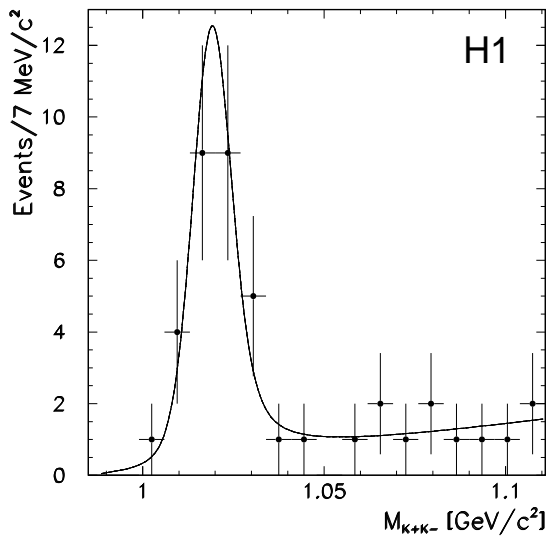,width=8cm,height=8cm}
\end{center}
\vspace{-0.5cm}
\caption{\mkk~invariant mass distribution for the selected events; the curve
is the result of a fit to a Breit-Wigner distribution convoluted with a Gaussian,
over a linear background.}
\label{fig:mass_kk}
\end{figure}

Fig.~\ref{fig:mass_kk} shows the invariant $K^+K^-$ mass distribution 
for the events passing all selection criteria, the kaon mass having
been assigned 
to the detected particles in the central tracking detector. 
A clear \ph\ signal over little background is observed in the mass region 
$1.00 < \mkk < 1.04$ \gevcsq, which contains 29 events.
%
%==============================================================================
%==============================================================================
The curve superimposed on Fig. \ref{fig:mass_kk} is the result of a fit to a  
relativistic Breit-Wigner distribution with a width fixed at the 
nominal value \cite{PDG}, convoluted
with a Gaussian distribution, over a linear background starting at threshold
($M = 2~M_K$). 
The r.m.s. width of the Gaussian distribution, which reflects 
the detector resolution, 
is fixed at a value of 4.5 \mevcsq\ as
obtained from MC.
Here, as in the \rh\ analysis, 
the program DIFFVM \cite{Benno} is used for MC simulation.
The fitted mass of the \ph\ meson is $1019 \pm 2$ \mevcsq,
in good agreement with the nominal value of $1019.4$ \mevcsq.

\hspace*{-8pt}The non-\ph\ background under the signal in the mass region 
$1.00<\mkk<1.04~\gevcsq$ is estimated to be $6 \pm 4 \%$. 
The error includes statistical and systematic uncertainties estimated by 
varying the shape of the background and the limits of the fit.
Because of the limited statistics, which do not allow an independent estimate, 
the background due to diffractive \ph\ production with proton dissociation 
is taken to be the same as for the \rh\ analysis, i.e. $9 \pm 8 {\%}$ of the 
selected signal, as would be expected with a factorisation of the diffractive
vertex.

After background subtraction and corrections for efficiencies and acceptances,
%determined from the Monte Carlo simulation and cross checked on the data,
for QED radiation effects and for the known \ph\ decay branching ratio into a 
$K^+K^-$~pair, the $ep$ cross section for elastic \ph\ production is
$$
 \sigma(e p \rightarrow e \ph\ p) = 
               50.7 \pm 11.8\ {\rm (stat.)} \pm 6.4\ {\rm (syst.)\ pb,}
$$ 
integrated over the range $6 < \qsq\ < 20$ \gevsq\ and
$42 < W < 134$ \gev.

The $ep$ cross sections are converted into \gsp\ cross sections using
a relation similar to eq.~(\ref{eq:sigma}).
Uncertainties in the \qsq\ and \W\ dependences measured in the present data
lead to systematic errors included in the quoted results.

The \gsp\ cross section for \ph\ meson elastic production, measured at 
$ \av{\W}$ $\approx 100$ \gev, is
$$
 \sigma(\gamma^* p \to \ph\ p) =  9.6 \pm 2.4\ {\rm nb}\ \ \ {\rm at} \  
      \av{\qsq} = 8.3\ \gevsq ,
$$
$$ \sigma(\gamma^ *p \to \ph\ p) =  3.1 \pm 1.0\ {\rm nb}\ \ \ {\rm at} \
      \av{\qsq} = 14.6\ \gevsq. 
$$
The quoted errors are the quadratic sums of the statistical and the systematic
uncertainties.

\begin{figure}[htbp]
\vspace{-1.1cm}
\begin{center}
\epsfig{file=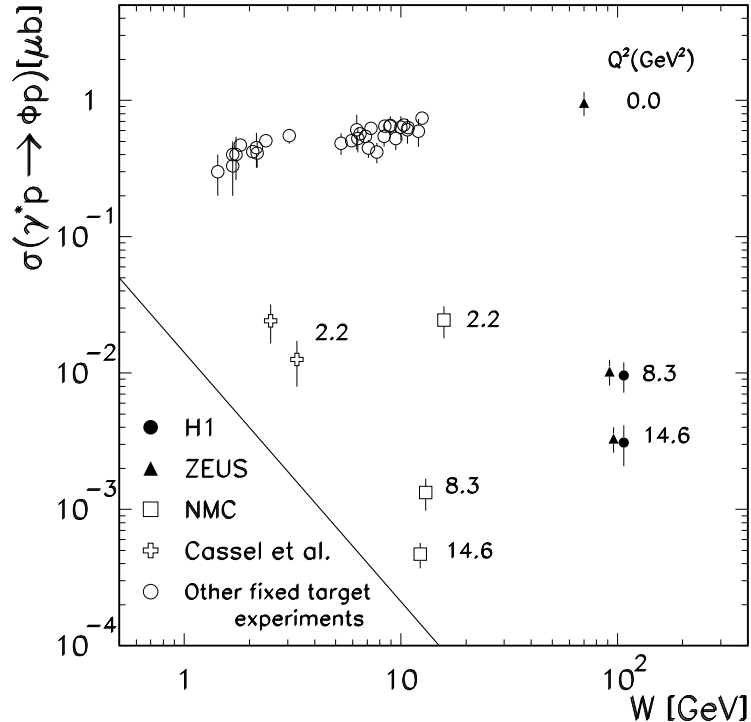,width=11cm,height=11cm}
\end{center}
\vspace{-0.5cm}
\caption{Cross section for $\gsp\ \to \ph p$ as a function of \W\
for several values of $\qsq$~($\gevsq$).
The overall normalisation uncertainties of $25\%$ for 
the results of Cassel et al.,
$20\%$ for NMC and $32\%$ for ZEUS are not included. The H1 errors are the
quadratic sums of the statistical and the systematic uncertainties.}
\label{fig:w_dep}
\end{figure}

These results are presented in Fig.~\ref{fig:w_dep}, 
together with a compilation of photoproduction and leptoproduction results
\cite{zphip,zphie,nmcrhophie,Bauer,fixedtgtphiphot,cassel,alekhin}
as a function of $W$.  
The NMC measurements were scaled to the values \qsq $=$ 2.1, 8.3 and 14.6 \gevsq\
using their measured \qsq\ dependence and the relevant values of the polarisation 
parameter $\varepsilon$; the ZEUS measurements are made at \qsq\ values very
close to ours. 
The overall normalisation uncertainties of 25 \% for the results of Cassel et al. 
\cite{cassel}, 20 \% for NMC \cite{nmcrhophie} 
and 32 \% for ZEUS  \cite{zphie} are not shown on the plot.

The cross section for elastic photoproduction of \ph\ mesons shows only a slow rise
from the fixed target to the HERA energies, consistent with soft pomeron exchange.
In contrast, at higher \qsq, the HERA values of the cross sections
are significantly larger than those of the NMC experiment, 
although the  errors are large and the comparison involves two
different experiments. 
% The change of behaviour of the cross section 
% with \qsq\ and \W\ appears faster for \ph\ than for \rh\ mesons 
% (compare with \cite{hele}). 

%
%==============================================================================

%==============================================================================

%==============================================================================
\subsection{{\boldmath \qsq} and {\boldmath $P_t^2$} Dependences and Polarisation}

\begin{figure}[htbp]
\vspace{-1.1cm}
\begin{center}
\epsfig{file=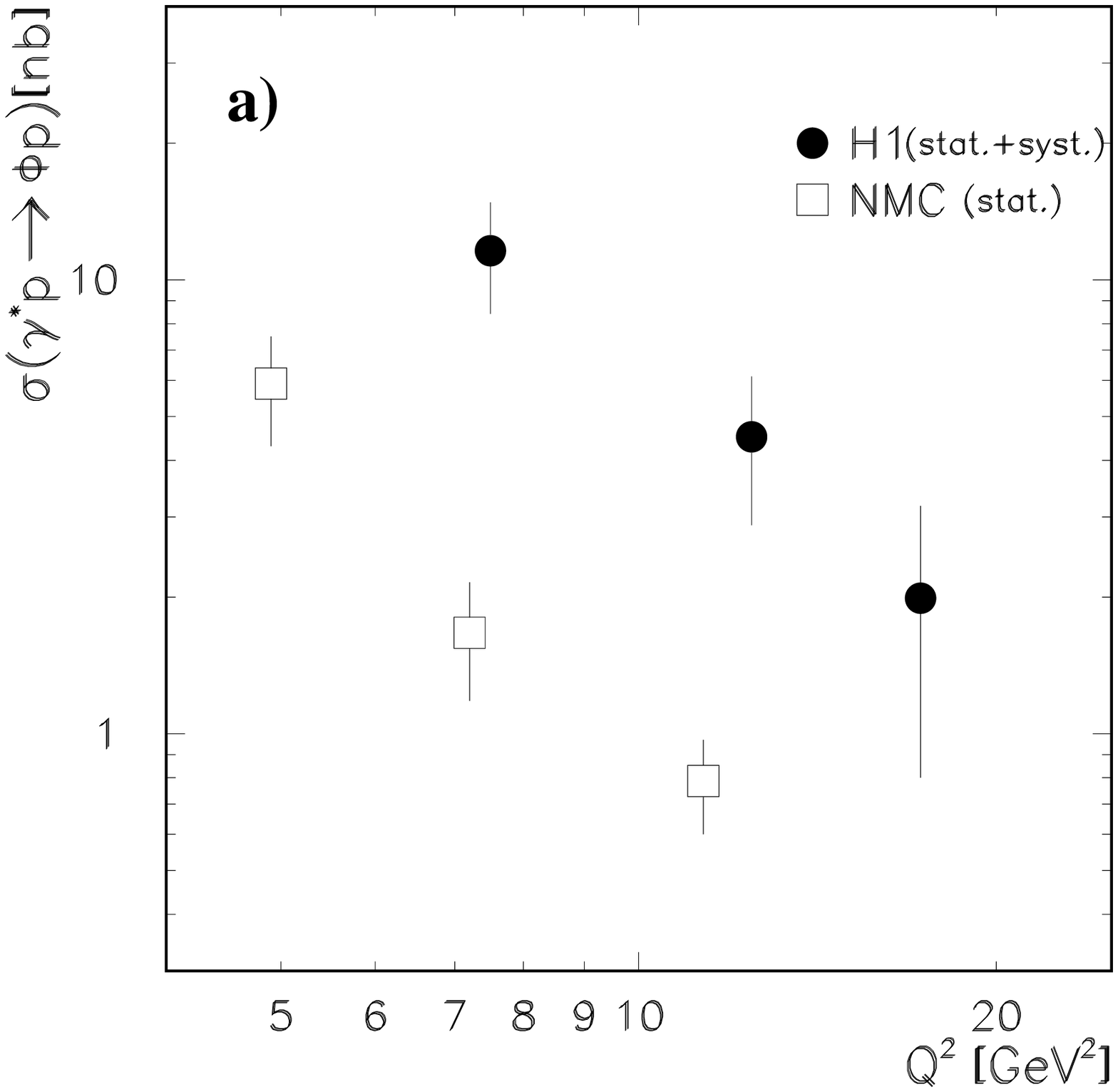,width=8cm,height=8cm}\epsfig{file=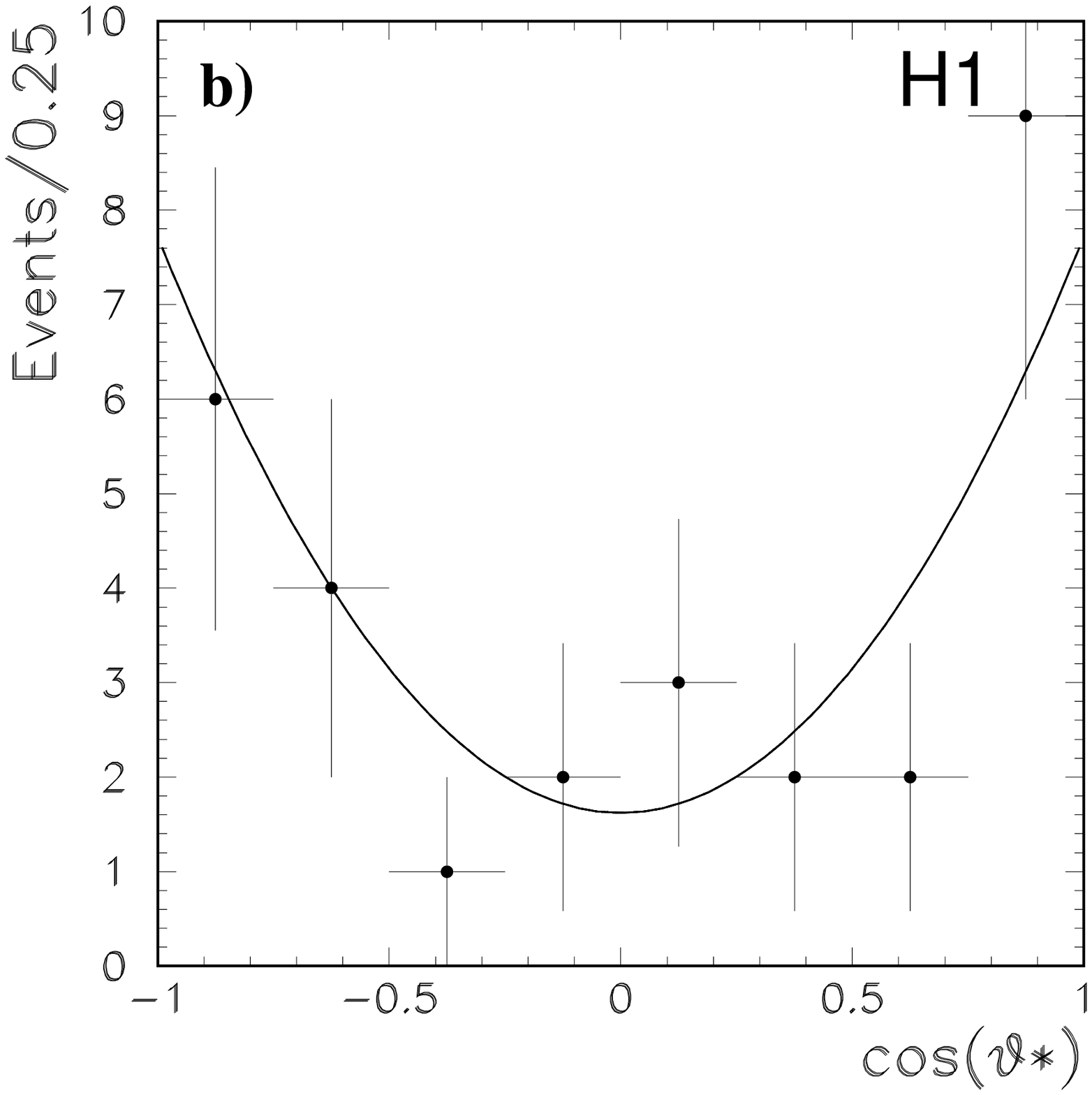,width=8cm,height=8cm}
\end{center}
\vspace{-0.5cm}
\caption{a) $Q^2$ dependence of the ${\gamma}^*p \to {\phi}p$ cross section.
b) Distribution of $\cos\vartheta^*$ for the selected \ph\ events;
the backgrounds are taken into account in the fit.
Only the statistical errors are shown.}
\label{fig:q2_dep}
\end{figure}

The \qsq\ dependence of the total \gsp\ cross section for elastic \ph\ 
production (see Fig.~\ref{fig:q2_dep}a) can be described by the form 
$Q^{-n}$ with $n=4.0 \pm 1.5 \pm 0.6 $,
where the first error is statistical and the second reflects the uncertainty on the
background contribution and the spread of results according to the 
details of the fitting procedure. 
%The non-resonant background was corrected for using its \qsq\ dependence 
%obtained from the data.
This value is close to that obtained by the NMC Collaboration~\cite{nmcrhophie}, 
with $n = 4.5 \pm 0.8 $, and by the ZEUS Collaboration~\cite{zphie}, with
$n = 4.1 \pm 1.2\ {\rm (stat.)}$.
It is also similar to that measured for the \rh\ 
($n = 5.0 \pm 1.0 \pm 0.4$ \cite{hele}).
% , and with the modified predictions based 
% on a perturbative QCD approach \cite{F_K_S,Nemch}.

%
%==============================================================================
%
%==============================================================================

An exponential fit to the $P_t^2$ distribution for $P_t^2 < 0.6$ \gevsq\ gives for
the slope the value 
$b = 5.2 \pm 1.6\ {\rm (stat.)} \pm 1.0\ {\rm (syst.)}\ \gevsqm $.
% (see Fig. \ref{fig:t_dep}).
The result is corrected for the presence of the two backgrounds mentioned
above, with fixed relative contributions with respect to the elastic signal 
and with fixed slopes.
The slope of the proton dissociation background is assumed to be 
$b = 2.0 \pm 1.0 $ \gevsqm\
(no dependence of the slope on the proton excitation mass is assumed) and
the slope for the non-resonant background is taken \linebreak to be 
$b = 0.4 \pm 0.4\, \gevsqm$, as obtained from 
a fit to the $P_t^2$ distribution of the events \linebreak with
$1.05 < \mkk < 1.3\, \gevcsq$ and which do not belong to the \rh\ peak 
\linebreak ($\mpipi < 0.4\, \gevcsq$).
The systematic error includes the effects of varying the amount of each
background and its slope within uncertainties, and those of changing the
details of the fitting procedure.

As for the \rh, the acceptance corrected distribution of $\cos \vartheta^*$
allows the extraction of the parameter \rzero\ (see Fig.~\ref{fig:q2_dep}b).
After subtraction of the non-resonant background, which 
is consistent with being flat in $\cos \vartheta^*$, and 
correction for detector effects, the fit gives 
$\rzero = 0.77 \pm 0.13 \pm 0.02 $, where 
the first error is statistical, and the second reflects the uncertainty on the 
background subtraction. 
This result is close to that obtained for elastic and proton
dissociative \rh\ production. Thus elastically electroproduced
\ph\ mesons are observed 
to be mostly longitudinally polarised, in agreement 
with model \linebreak predictions \cite{Cudell,F_K_S}.

%==============================================================================
%
\subsection{Cross Section Ratio {\boldmath ${\sigma}({\phi})/{\sigma}({\rho})$} }

The study of the \qsq\ and \W\ evolutions
of the $\ph\ / \rh$ cross section ratio for elastic production is of 
particular interest. Systematic uncertainties largely cancel in the ratio
as the same selection criteria are used for the two event samples.
It is thus possible to relax the requirements on the 
positron cluster position and energy deposition in the BEMC, and those 
on the associated BPC hit, since the trigger conditions and the kinematic 
variable reconstruction are affected in the same way for both samples.
The accepted kinematic domain is then extended to $5 <\qsq < 20$ \gevsq.
A slight difference in the acceptances for the two samples, due to 
the different opening angles in the laboratory system between 
the two decay hadrons, is taken into account.

\begin{figure}[htbp]
\vspace{-1.cm}
\begin{center}
\epsfig{file=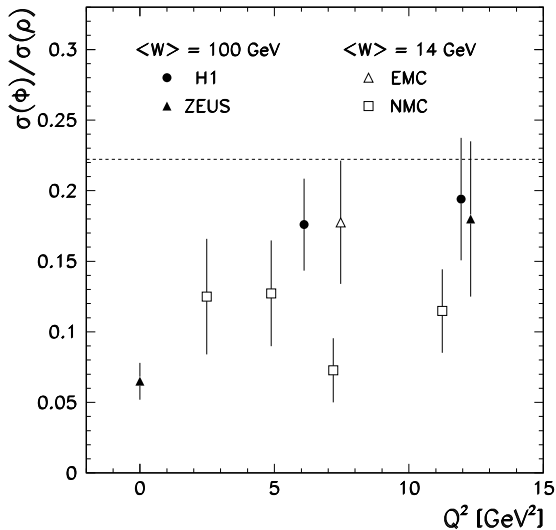,width=8cm,height=8cm}
\end{center}
\vspace{-0.5cm}
\caption{Ratio of the cross sections for elastic \ph\ and \rh\ production,
as a function of \qsq. 
The H1 and ZEUS (at $\qsq \approx 12.3~\gevsq$) errors are 
the quadratic sums of the
statistical and systematic uncertainties. For the EMC, NMC
and ZEUS (at $\qsq \approx 0~\gevsq$) points only statistical errors
are shown. The overall normalisation uncertainties of about
10$\,\%$ for the NMC and EMC results are not included.
The dashed line corresponds to the ratio $2/9$
from quark charge counting and SU(3).}
\label{fig:ratio}
\end{figure}

After acceptance corrections and background subtraction, 
the \ph\ to \rh\ cross section ratio $R_{\ph / \rh}$ is measured to be 
$$
  R_{\ph / \rh} =  0.18 \pm 0.03\ \ \ {\rm at} \  \av{\qsq} = 6.1\ \gevsq
\ \ (5 <\qsq < 8.3\ \gevsq) ,
$$
$$
  R_{\ph / \rh} =  0.19 \pm 0.04\ \ \ {\rm at} \  \av{\qsq} = 12.0\ \gevsq
\ \ (8.3 <\qsq < 20\ \gevsq),
$$
for \av{\W} $\approx 100$ \gev.
These results are presented in 
Fig.~\ref{fig:ratio}, together with
the high \qsq\ results of the EMC \cite{emcphie}, NMC \cite{nmcrhophie} and ZEUS 
Collaborations \cite{zphie}, and with the ZEUS photoproduction
measurement \cite{zphip}. 
%
%
%==============================================================================
%
%==============================================================================

%The present result confirms the significant rise of the 
%cross section ratio with \qsq. 
The cross section ratio values from HERA at high \qsq\ are close
to the SU(3) quark charge counting prediction 
of $2 / 9$, in contrast with photoproduction results
($R_{\ph / \rh} \simeq 0.07$). \linebreak
Approaches based on perturbative QCD predict
that this value for the ratio should be exceeded at very large \qsq\ 
\cite{F_K_S,Nemch}.
In Fig.~\ref{fig:ratio} there is 
also an indication that, for comparable \qsq\ values, the ratio $R_{\ph / \rh}$
may be higher at HERA than for the fixed target experiments.

%=======================================================================
\section{Summary} \label{sect:summary}

The electroproduction of $\rho$ mesons with proton
diffractive dissociation has been studied at HERA with
the H1 detector.
The ratio of the proton 
dissociative to the elastic cross section is measured 
to be $\sigma ( \gamma^* p \rightarrow \rho Y )/
{\sigma ( \gamma^* p \rightarrow \rho p )}=0.65 \pm 0.11 \pm 0.13$,
in the kinematic range $7 < Q^2 <  35\, \GeVsq$ and $60 < W < 180\, \GeV$
(corresponding to \av{\qsq} $\approx 10\, $ \gevsq\ and
\av{\W} $\approx 128\, $ \gev ) for the mass interval $M_Y^2 / W^2 < 0.05$.
% at \av{\W} $\approx 128\, $ \gev\ and \av{\qsq} $\approx 10\, $ \gevsq.
Dividing the kinematic range into two $W$ and two $\Qsq$ intervals
gives no significant indication of variation of this ratio.

The exponential slope of the $P_t^2$ distribution for 
proton dissociative $\rho$ meson electroproduction is
found to be $b_{pdis} =2.1 \pm 0.5 \pm 0.5$ GeV$^{-2}$,
with no significant $M_Y$ dependence.
The $\Qsq$ distribution for the $\gamma^* p \rightarrow \rho Y$
process is well fitted by the form $Q^{-n}$,
with $n = 5.8 \pm 1.1 \pm 0.8$, similar to the value
$n = 5.0 \pm 1.0 \pm 0.4$ for elastic $\rho$ production.
The probability $r^{04}_{00}$ for the $\rho$ meson to be 
longitudinally 
polarised is measured to be $r^{04}_{00} = 0.79 \pm 0.10 \pm 0.05$,
close to the value $r^{04}_{00}= 0.73 \pm 0.05 \pm 0.02$
for elastic $\rho$ production,
demonstrating that $\rho$ mesons are mostly longitudinally polarised
also in the proton dissociative process.

The similarity of the $\Qsq$ and polarisation behaviour for
$\rho$ meson electroproduction in proton dissociative and in elastic 
scattering, and the comparison of these results with
proton-proton collider results, do not provide any evidence 
of correlations between the processes originating 
from different vertices, i.e. for breaking of factorisation.

The cross section for the elastic electroproduction of \ph\ mesons has been
measured at HERA in the kinematic range $6 < \qsq < 20$ \gevsq\ 
and $42 < \W < 134$ \gev.
At this \qsq\ the cross section is significantly larger than observed in the
fixed target measurement \cite{nmcrhophie} at smaller \W, in contrast
to the photoproduction case.
The \qsq\ dependence of the \gsp\ cross section is well described by the
form $Q^{-n}$ with $n=4.0 \pm 1.5 \pm 0.3$. The exponential slope
of the $P_t^2$ distribution 
is found to be $b = 5.2 \pm 1.6 \pm 1.0\ \gevsqm$.
Elastically electroproduced \ph\ mesons are found 
to be mostly longitudinally polarised.

The ratio of the cross sections for elastic electroproduction 
of \ph\ and \rh\ mesons is measured, for $5 <\qsq < 20$ \gevsq, to be
$R_{\ph / \rh} =  0.18 \pm 0.03$.
This ratio is significantly larger than in photoproduction, and the
comparison with fixed target results provides some indication that,
for comparable \qsq\ values, the ratio is may be higher 
at HERA than at lower energy.

%=======================================================================
%=============================================================
\section*{Acknowledgments}

We are grateful to the HERA machine group whose outstanding
efforts have made and continue to make this experiment possible. We thank
the engineers and technicians for their work in constructing and now
maintaining the H1 detector, our funding agencies for financial support, the
DESY technical staff for continual assistance, and the DESY directorate for the
hospitality which they extend to the non--DESY members of the collaboration.
We further thank A.B. Kaidalov, L.L. Frankfurt and E.M. Levin
for useful discussions.

%=======================================================================
%=============================================================
%\begin{thebibliography}{19}

%


\begin{thebibliography} {99}

%
\bibitem {help}
  S.\ Aid et al., H1 Coll., \np {B463} {1996} {3}.            % H1 photoprod rho 
%
\bibitem {hjpp}
  S.\ Aid et al., H1 Coll., \np {B472} {1996} {3}.            % H1 photoprod jpsi
%
\bibitem {zrhop}                                              % Zeus photoprod
  M.\ Derrick et al., ZEUS Coll., \zp {C69} {1995} {39}; \\         %rho
  M.\ Derrick et al., ZEUS Coll., preprint DESY-96-183 (1996).      %rho LPS
%
\bibitem {zomep}
  M.\ Derrick et al., ZEUS Coll., \zp {C73} {1996} {73}.   %Zeus photoprod omega
%
\bibitem {zphip}
  M.\ Derrick et al., ZEUS Coll., \pl {B377} {1996} {259}.     %Zeus photoprod  phi
%
\bibitem {zjpsip}
  M.\ Derrick et al., ZEUS Coll., \pl {B350} {1995} {120}.      %Zeus photoprod jpsi
%
\bibitem {hele}
   S.\ Aid et al., H1 Coll., \np {B468} {1996} {3}.           % H1 electroprod rho+jpsi
%
\bibitem{zrhoe}                                                 %zeus electroprod rho
  M.\ Derrick et al., ZEUS Coll., \pl {B356} {1995} {601}.       
%
\bibitem{zphie}                                                 %zeus electroprod phi
  M.\ Derrick et al., ZEUS Coll., \pl {B380} {1996} {220}.           
%
\bibitem {gp_rho}
  R.M.\ Egloff et al., \prl {43} {1979} {657}; \\             %  photoprod rho
  D.\ Aston et al., \np {B209} {1982} {56}.
%
\bibitem {gp_jpsi} 
  U.\ Camerini et al., \prl {35} {1975} {483; \\    %SLAc - real photons.
  M.\ Binkley et al., \prl {48} {1982} {73}; \\     %E401 
  B.H.\ Denby et al., \prl {52} {1984} {795}; \\    %E516 - FTPS
  R.\ Barate et al., \zp {C33} {1987} {505}; \\     %NA14
  P.L.\ Frabetti et al., \pl {B316} {1993} {197}.   %E687
%
\bibitem {nmcrhophie}
  M.\ Arneodo et al., NMC Coll., \np {B429} {1994} {503}.       %nmc muoprod rho + phi
%
\bibitem {emcphie}
  J.\ Ashman et al., EMC Coll., \zp {C39} {1988} {169}.         %emc muoprod phi
%
%\bibitem {emcrhoe}
%  J.J.\ Aubert et al., EMC Coll., \np {B213} {1983} {1}.
%
\bibitem {cdpp}
  F.\ Abe et al., CDF Coll., \prev {D50} {1994} {5518}; \\
  F.\ Abe et al., CDF Coll., \prev {D50} {1994} {5535}.
%
\bibitem {uafpp}
  M.\ Bozzo et al., UA4 Coll., \pl {B147} {1984} {392}; \\
  D.\ Bernard et al., UA4 Coll., \pl {B186} {1987} {227}.
%
\bibitem {uappp}
  R.E.\ Ansorge et al., UA5 Coll., \zp {C33} {1986} {175}.
%
\bibitem {isrpp}
  J.C.M.\ Armitage et al., CHLM Coll., \np {B194} {1982} {365}; \\
  N.\ Amos et al., \pl {B120} {1983} {460}.
%  J.C.M.\ Armitage et al., CHM Coll., \np {B132} {1978} {365}. 
%
\bibitem {regge}
  P.D.B.\ Collins, {\it An Introduction to Regge Theory and High-Energy 
  Physics}, Cambridge University Press, Cambridge, England, 1977.
%  D.P.\ Roy and R.G.\ Roberts, \np {B77} {1974} {240}; \\
%  R.D.\ Field and G.C.\ Fox, \np {B80} {1974} {367}.
%
\bibitem {VDM} 
  J.J.\ Sakurai, \prl {22} {1969} {981}; \\
  J.J.\ Sakurai and D.\ Schildknecht, \pl {B40} {1972} {121}.
%
\bibitem {Bauer}
  T.H.\ Bauer et al., \rmp {50} {1978} {261}, and references therein.
%
\bibitem  {Low_Nussinov} 
  F.E.\ Low, \prev {D12} {1975} {163};\\
  S.\ Nussinov, \prl {34} {1975} {1268}.
%
\bibitem {DL}
  A.\ Donnachie and P.V.\ Landshoff, \pl {B348} {1995} {213}.
%
\bibitem {Cudell} 
  J.R. \ Cudell, \np {B336} {1990} {1}.
%
\bibitem {Ryskin} 
  M.G.\ Ryskin, \zp {C57} {1993} {89}.
%
\bibitem {Zakha}
  B.Z.\ Kopeliovich et al., \pl {B324} {1994} {469}; \\
  N.N.\ Nikolaev, B.G.\ Zakharov and V.R.\ Zoller \pl {B366} {1996} {337}; \\
  J.\ Nemchik et al., \pl {B374} {1996} {199}.
%
\bibitem {Haak} 
  L.P.A.\ Haakman, A.\ Kaidalov and J.H.\ Koch, \pl {B365} {1996} {411}.
%
\bibitem {Ginzburg} 
  I.F.\ Ginzburg and D.Yu.\ Ivanov, preprint hep-ph/9604437 (1996).
%
\bibitem {Brodsky} 
  S.J.\ Brodsky et al., \prev {D50} {1994} {3134}.
%
\bibitem {Levin}
E.\ Gotsman, E.M.\ Levin and U.\ Maor, \np {B464} {1996} {251}.
%
\bibitem {F_K_S} 
  L.\ Frankfurt, W.\ Koepf and M.\ Strikman, \prev {D54} {1996} {3194}.
%
\bibitem {Nemch} 
  J.\ Nemchick et al., 
    preprint DFTT 71/95, KFA-IKP(TH)-24-95,hep-ph/9605231 (1996).
%    {\it Colour dipole phenomenology of diffractive 
%    electroproduction of light vector mesons at HERA}, 
%
\bibitem {h1det}
  I.\ Abt et al., H1 Coll.,  
     {\it The H1 detector at HERA,}
     \nim {A386} {1997} {310, 348}.
%     preprint DESY-93-103 (1993) and DESY internal report H1-96-01 (1996).
%
% \bibitem {fwdet}
%  T.\ Ahmed et al., H1 Coll., \pl {B348} {1995} {681}.
%
%\bibitem  {Kope} 
%  B.Z. \ Kopeliovich and B.G. Zakharov, \prev {D44} {1991} {3466};\\
%  B.Z.\ Kopeliovich et al., \pl {B324} {1994} {469};\\
%  J.\ Nemchik et al., \pl {B341} {1994} {228}.
%
%
%
\bibitem {doan}
  S.\ Bentvelsen, J.\ Engelen and P.\ Kooijman, 
    Proc. of the Workshop on Physics at HERA, 
    ed. W. Buchm\"uller and G. Ingelman, Hamburg 1992, Vol. 1, p. 23; \\
  K.C.\ Hoeger, ibid., p.43; \\
  G.\ Wolf, preprint DESY-94-022 (1994).
%
\bibitem {Kaidalov}
 A.B.\ Kaidalov, \prep {50} {1979} {157}.
%
\bibitem {Goulianos}
  K.\ Goulianos, \prep {101} {1983} {169}.
%
\bibitem {factor}
  H.\ Holtmann at al., \zp {C69} {1996} {297}.
%
% \bibitem {Schilling_Wolf}
%   K.\ Schilling and G.\ Wolf, \np {B61} {1973} {381}.
%
% \bibitem {RS}
%  M.\ Ross and L.\ Stodolsky, \prev {149} {1966} {1172}.
%
\bibitem {IDL}
A.\ Donnachie and P.V.\ Landshoff, \np {B231} {1984} {189}.
%
\bibitem {ftest}
 D.S.\ Ayres et al., \prl {37} {1976} {1724}.
%
\bibitem {qmod}
  S.P.\ Misra, A.R.\ Panda and B.K.\ Parida, \prev {D22} {1980} {1574}.
%
\bibitem {isrdd}
 C.\ Conta et al., \np {B175} {1980} {97}.
%
\bibitem {Benno}
 B.\ List, {\it DIFFVM program,}
    {\it Diploma Thesis}, Techn. Univ. Berlin, unpubl. (1993).
%
\bibitem {Heracles}
  H.\ Spiesberger,
    {\it HERACLES} 4.4,
    unpublished program manual (1993); \\
  A.\ Kwiatkowski, H.\ Spiesberger and H.-J.\ M\"ohring, 
    Proc. of the Workshop on Physics at HERA, 
    ed. W. Buchm\"uller and G. Ingelman, Hamburg 1992, Vol. 3, p. 1294.
%
\bibitem {Jackson}
  J.D.\ Jackson, \nc {34} {1964} {1644}.
%
\bibitem {PDG}
 R.M.\ Barnett et al., 
   {\it Review of Particle Physics},
   \prev {D54} {1996} {1}.

%
\bibitem {DoLa} 
  A.\ Donnachie and P.V.\ Landshoff, \pl {B296} {1992} {227}.
%
\bibitem {ozi} 
  S.\ Okubo, \pl {5} {1963} {165}; \\
  G.\ Zweig, CERN Reports TH-401, TH-412 (1964); \\
  J.\ Iizuka, {\em Prog. Theor. Phys. Suppl.} {\bf 37/38} (1966) 21.
%
%
\bibitem {fixedtgtphiphot}
  R.M.\ Egloff et al., \prl {43} {1979} {657};\\
  J.\ Busenitz et al., \prev {D40} {1989} {1}.
%
%\bibitem {DoLaVM} 
%  A.\ Donnachie and P.V.\ Landshoff, \pl {B348} {1995} {213}. - check
%
\bibitem {cassel}
  \ D.G.\ Cassel et al., \prev {D24} {1981} {2787}.
%
\bibitem {shambroom}
  \ W.D.\ Shambroom et al., CHIO Coll., \prev {D26} {1982} {1}.
%

\bibitem {alekhin} 
  S.I.\ Alekhin et al., 
    CERN-HERA 87-01(1987), ed. H.\ Schopper in: Landolt-Bornstein, New Series, Vol. 12b.
%    {\it Total Cross-Sections for Reactions of High Energy Particles},
%
}
\end{thebibliography}
\end{document}